\documentclass[11pt]{article}
\usepackage{graphicx}
\usepackage{epsfig}
\usepackage{amsfonts,amssymb}
\usepackage{latexsym}
\usepackage{cite}
\newcommand{\dd}{{\rm d}}

\newcommand{\eq}{\begin{equation}}
\newcommand{\feq}{\end{equation}}
\newcommand{\eeq}{\end{equation}}
\newcommand{\eqn}{\begin{eqnarray}}
\newcommand{\feqn}{\end{eqnarray}}
\newcommand{\arr}{\begin{eqnarray*}}
\newcommand{\farr}{\end{eqnarray*}}

\newcommand{\M}{{\cal M}}
\newcommand{\calP}{{\cal P}}

\newcommand{\calS}{{\cal S}}

\newcommand{\DD}{{\cal D}}
\newcommand{\VV}{{\cal V}}

\newcommand{\F}{{\cal F}}

\newcommand{\T}{{\cal T}}

\newcommand{\ie}{{\it i.e.,}\ }
\newcommand{\p}{\partial}

\newcommand{\lp}{\left(}
\newcommand{\rp}{\right)}
\font\mybb=msbm10 at 12pt
\def\bb#1{\hbox{\mybb#1}}

\def\bR {\bb{R}}

\def\ga{\gamma}

\def\ka{\kappa}

\def\si{\sigma}
\def\om{\omega}

\def\th{\theta}

\def\Si{\Sigma}

\newcommand{\beq}{\begin{equation}}
\newcommand{\beqa}{\begin{eqnarray}}
\newcommand{\eeqa}{\end{eqnarray}}

\begin{document}

\setlength{\unitlength}{1mm}

\thispagestyle{empty}
%\rightline{\small hep-th/08}
 \vspace*{0.5cm}

\begin{center}
{\bf \LARGE Dyonic AdS black holes from magneto\-hydro\-dynamics}\\

\vspace*{2cm}

{\bf Marco M.~Caldarelli,}$^{1,2}\,$ {\bf \'Oscar
J.~C.~Dias}$^{1,3}\,$ and {\bf Dietmar Klemm}$^4\,$

\vspace*{0.5cm}

{\it $^1\,$Departament de F{\'\i}sica Fonamental, Universitat de
Barcelona, \\
Marti i Franqu{\`e}s 1,
E-08028 Barcelona}\\[.3em]
{\it $^2\,$Instituut voor Theoretische Fysica, Katholieke Universiteit Leuven, \\
Celestijnenlaan 200D B-3001 Leuven, Belgium}\\[.3em]
{\it $^3\,$Dept. de F\'{\i}sica e Centro de F\'{\i}sica do Porto,
Faculdade de Ci\^encias da Universidade do Porto, Rua do Campo
Alegre 687, 4169 - 007 Porto, Portugal}\\[.3em]
{\it $^4\,$Dipartimento di Fisica dell'Universit\`a di Milano, \\
Via Celoria 16, I-20133 Milano and \\
INFN, Sezione di Milano, Via Celoria 16, I-20133 Milano.}\\[.3em]

\vspace*{0.3cm} {\tt marco.caldarelli@fys.kuleuven.be, oscar.dias@fc.up.pt,
dietmar.klemm@mi.infn.it}

\vspace*{2cm}

\vspace{.8cm} {\bf ABSTRACT}
\end{center}

We use the AdS/CFT correspondence to argue that large dyonic black holes
in anti-de~Sitter spacetime are dual to stationary solutions of the equations of
relativistic magnetohydrodynamics on the conformal boundary of AdS. The dyonic
Kerr-Newman-AdS$_4$ solution corresponds to a charged diamagnetic fluid not
subject to any net Lorentz force, due to orthogonal magnetic and electric fields
compensating each other. The conserved charges, stress tensor and R-current
of the fluid are shown to be in exact agreement with the corresponding
quantities of the black hole. Furthermore, we obtain stationary solutions of
the Navier-Stokes equations in four dimensions, which yield
predictions for (yet to be constructed) charged rotating black strings in
AdS$_5$ carrying nonvanishing momentum along the string. Finally, we
consider Scherk-Schwarz reduced AdS gravity on a circle. In this theory,
large black holes and black strings are dual to lumps of deconfined
plasma of the associated CFT. We analyze the effects that a magnetic
field introduces in the Rayleigh-Plateau instability of a plasma
tube, which is holographically dual to the Gregory-Laflamme instability of a
magnetically charged black string.

\noindent

%\keywords{AdS-CFT correspondence, black holes}

\vfill \setcounter{page}{0} \setcounter{footnote}{0}
\newpage

\tableofcontents

%%%%%%%%%%%%%%%%%%%%%%%%%%%%%%%%%%%%%%%%%%%%%%%%%%%%%%%%%%%%%%%%%%%%%%%%%%%%%
\setcounter{equation}{0}\section{Introduction}

The AdS/CFT correspondence (cf.~\cite{Aharony:1999ti} for a review)
has provided us with a powerful tool to get insight into the strong coupling
dynamics of certain field theories by studying classical gravity.
In the long wavelength limit, it is reasonable to expect that these
interacting field theories admit an effective description in terms of
hydrodynamics. One can thus conversely use the equations of fluid mechanics
(which, in simple contexts, are much easier to solve than the
full set of Einstein's equations) in order to make predictions
for the gravity side. This was done in \cite{Lahiri:2007ae}, where the
fluid configurations dual to (yet to be discovered) black rings in
(Scherk-Schwarz compactified) AdS were constructed\footnote{For a discussion
of plasma lumps dual to black saturns in AdS cf.~\cite{Evslin:2008py}.}.

More recently, it was shown in \cite{Bhattacharyya:2008jc}
(cf.~also \cite{VanRaamsdonk:2008fp} for the
four-dimensional case) that the equations of hydrodynamics (i.~e., the
Navier-Stokes equations) can also be derived directly from Einstein's
equations with negative cosmological constant, without making any use of
string theory, from which these ideas originally emerged.

The correspondence between gravity in asymptotically AdS spaces and
fluid mechanics has interesting consequences: For instance, it is
well known that under certain conditions fluids are affected by the
so-called Rayleigh-Plateau instability (responsible e.~g.~for the
pinch-off of thin water jets from kitchen taps), and one might ask
what the gravity dual of (the relativistic analogue of) this
instability is. This point was studied recently in
\cite{Caldarelli:2008mv}, where it was argued that the gravity dual
is the Gregory-Laflamme instability \cite{Gregory:1993vy}.

In this paper, we shall consider a more general setting in which the charged
fluid moves in external magnetic fields. The fluid is thus described by
the equations of magnetohydrodynamics (MHD). Such a generalization is
of interest for several reasons:
First, there are two possible fall-off conditions for a bulk U(1) gauge field:
A normalizable mode that corresponds to a VEV of the dual operator
(an R-current), and a non-normalizable one corresponding to the application
of an external gauge field, that can be thought of as an electromagnetic
field. (2+1)-dimensional field theories deformed in this way, that are dual
to magnetically charged AdS$_4$ black holes, have become fashionable recently
in the context of a possible holographic realization of condensed matter
phenomena like superconductivity \cite{Hartnoll:2008vx,Gubser:2008px}, the Hall
effect \cite{Hartnoll:2007ai} and the Nernst effect \cite{Hartnoll:2007ih}.
In the long wavelength regime a quantum field theory in presence
of external gauge fields is expected to be described by MHD. Second, solving
the MHD equations (which, under certain symmetry assumptions, might be
much easier than solving the full set of Einstein-Maxwell equations) can be
helpful for constructing new magnetically
charged black hole solutions in AdS. Furthermore, the phase structure of
such AdS black holes can be studied in a simplified setting using
magnetohydrodynamics.

We should stress that the magnetic field entering the MHD equations is
nondynamical, so that there are no Maxwell equations on the
boundary. Note however that, in the same way in which one can
promote the CFT metric to a dynamical field \cite{Compere:2008us}, it should
be possible to add dynamics also to the magnetic field, although we will
not try to do this here.

The remainder of this paper is organized as follows:
In section \ref{sec:MHD}, we review general aspects of MHD, like
the equation of state and the stress tensor of the fluid under the
influence of external magnetic fields. Moreover,
we show how the MHD equations emerge from Einstein-Maxwell-AdS gravity.
In the following section, we generalize the results
of \cite{Bhattacharyya:2007vs} to the case of nonvanishing magnetic fields.
We first consider static dyonic black holes in
AdS$_4$. In an appropriate limit (when the horizon radius $r_h$ is much larger
than the AdS curvature radius $\ell$), these black holes are effectively
described by magnetohydrodynamics. Conformal invariance and extensivity
imply that the grandcanonical partition function has to take the form
\eq
\frac 1V\ln{\cal Z}_{gc} = T^2h(\zeta/T,B/T^2)\,, \label{part-fct-stat}
\feq
where $\zeta$ denotes a chemical potential conjugate to the U(1) electric
charge of the fluid, $B$ is the applied magnetic field, and $V$ and $T$
represent the volume and the overall temperature of the fluid respectively.
We then show that, using the function $h$ that can be read off from the
partition function of static black holes as an input into the MHD
equations, we can exactly reproduce the conserved charges, boundary
stress tensor and R-currents associated to the rotating dyonic
Kerr-Newman-AdS solutions. Their thermodynamics is very simple: It is
summarized by the partition function
\eq
\frac 1V\ln{\cal Z}_{gc} = \frac{T^2h(\zeta/T,B\Xi/T^2)}{\Xi}\,,
\feq
with $\Xi=1-\omega^2\ell^2$, and $\omega$ is the angular velocity of the
black hole.

After that, we solve the Navier-Stokes equations on $\bR\times S^1\times S^2$,
$\bR\times S^1\times\bR^2$ or $\bR\times S^1\times H^2$. This yields
predictions for (yet to be constructed) charged rotating black strings in
AdS$_5$ with momentum along the string.

In section \ref{sec:RPmag} we consider perturbations around plasma tube
solutions of the MHD equations on $\bR\times\bR^2\times S^1$, that are dual to
magnetic black strings on Scherk-Schwarz compactified AdS$_6$. These plasma
tubes suffer from the long wavelength Rayleigh-Plateau instability, which
is shown to be weakened by the presence of a magnetic background. This is
exactly what one expects from the known studies of the Gregory-Laflamme
instability for magnetically charged black strings in asymptotically flat
space, to which the considered SS-compactified AdS black strings are similar.

Throughout this paper we use calligraphic letters ${\cal T}, {\cal V}, {\cal B},
{\cal S},\ldots $ to indicate local thermodynamic quantities, whereas
$T,V,B,S,\ldots $ refer to the whole fluid configuration. $\mu$ and $\zeta$
are local and global chemical potentials respectively.\\

{\it Note added}: While this paper was in preparation, ref.~\cite{Hansen:2008tq}
appeared that partially overlaps with our section \ref{MHD-grav}.

%%%%%%%%%%%%%%%%%%%%%%%%%%%%%%%%%%%%%%%%%%%%%%%%%%%%%%%%%%%%%%%%%%%%%%%%%%%%%
\setcounter{equation}{0}\section{Magnetohydrodynamics\label{sec:MHD}}

\subsection{Equation of state\label{eqState}}

We would like to know how the equation of state ${\cal P}=\rho/(d-1)$
of a conformal fluid in $d$ dimensions changes in presence of a
magnetic field $\cal B$. To this end, consider the grandcanonical potential
\begin{equation}
\Phi = {\cal E} - {\cal T}{\cal S} - \mu{\cal R} = \Phi({\cal T},{\cal V},
\mu, {\cal B})\,. \label{grandcan-pot}
\end{equation}
It follows from conformal invariance and extensivity that
\begin{equation}
\Phi = -{\cal V}{\cal T}^d h(\nu, b)\,, \label{Phi}
\end{equation}
where we defined $\nu=\mu/{\cal T}$ and $b={\cal B}/{\cal T}^2$. Note that
$\mu$ and $\cal B$ have mass dimension one and two respectively, so that
$\nu$ and $b$ are dimensionless. Equation (\ref{Phi}) defines the function
$h(\nu,b)$. From (\ref{Phi}) it is also clear that
\begin{equation}
\Phi(\lambda{\cal T}, \lambda^{1-d}{\cal V}, \lambda\mu, \lambda^2 {\cal B})
= \lambda\Phi({\cal T},{\cal V},\mu, {\cal B})\,.
\end{equation}
If we derive this with respect to $\lambda$, set $\lambda=1$ and use
\begin{equation}
\frac{\partial\Phi}{\partial{\cal T}} = -{\cal S}\,, \qquad
\frac{\partial\Phi}{\partial{\cal V}} = -{\cal P}\,, \qquad
\frac{\partial\Phi}{\partial{\cal \mu}} = -{\cal R}\,, \qquad
\frac{\partial\Phi}{\partial {\cal B}} = -{\cal V}{\cal M}\,,
\end{equation}
where ${\cal M}$ is the magnetization density, we obtain
\begin{equation}
-{\cal S}{\cal T} - (1-d){\cal P}{\cal V} - {\cal R}\mu - 2{\cal V}{\cal M}
{\cal B} = \Phi\,.
\end{equation}
Using the definition (\ref{grandcan-pot}) of $\Phi$, this yields the
equation of state
\begin{equation}
\rho = (d-1){\cal P} - 2{\cal B}{\cal M}\,. \label{state-equ}
\end{equation}
We will see below that large dyonic black holes in AdS$_4$ indeed have a
grandcanonical potential of the form (\ref{Phi}), with $d=3$.

\subsection{Stress tensor of the fluid}
\label{stress-fluid}

We want to determine now the general form of the stress tensor for
our fluid, and we will do this in a derivative expansion in the
fluid velocity $u^\mu$, following \cite{Andersson:2006nr}. The zero
order term corresponds to a perfect fluid (under the influence of
external electromagnetic fields $F^I_{\mu\nu}$, labelled by the index $I$),
with stress tensor, charge currents and entropy current given respectively by
(cf.~e.~g.~\cite{Hartnoll:2007ih}) \eq T^{\mu\nu}=\rho u^\mu
u^\nu+{\cal P}\Pi^{\mu\nu} - \M_I^{\mu\lambda}
          {F^{I\nu}}_{\lambda}\,,\qquad J^\mu_I + \nabla_{\sigma}\M_I^{\sigma\mu}
          = r_Iu^\mu\,,\qquad J^\mu_S=su^\mu\,, \label{T-J-JS}
\eeq where we have introduced the projection tensor \eq
\Pi^{\mu\nu}=g^{\mu\nu}+u^\mu u^\nu \eeq on the directions
orthogonal to $u^\mu$. $\rho=\rho(\T,\mu^I,{\cal B}^J)$ is the rest frame
energy density, $r_I=r_I(\T,\mu^J,{\cal B}^K)$ and $s=s(\T,\mu^I,{\cal B}^J)$ are
the rest frame charge- and entropy densities, while $\mu^I$ and
${\cal B}^J$ denote the chemical potentials and magnetic fields
respectively. The polarization tensor $\M_I^{\mu\lambda}$ is defined
by
 \eq
\M_I^{\mu\lambda} = -\frac 1{\VV}\frac{\partial\Phi}{\partial
F^I_{\mu\lambda}}\,. \feq Note that the microscopic currents
(polarization currents) are given by \eq J^{\mu}_{I\,\mathrm{micr.}}
= -\nabla_{\sigma}\M_I^{\sigma\mu}\,, \feq whereas $J^{\mu}_I$
represents the total current density, so that the combination
$J^\mu_I + \nabla_{\sigma}\M_I^{\sigma\mu}$ appearing in
(\ref{T-J-JS}) is the macroscopic or transport current. Notice also
that the extra contribution $-\M_I^{\mu\lambda}{F^{I\nu}}_{\lambda}$
to the stress tensor in (\ref{T-J-JS}) comes from the coupling of
the polarization to the electromagnetic field.

The first subleading order in the derivative expansion of the MHD equations
describes dissipative phenomena like viscosity and diffusion. Lorentz
invariance and the requirement that the entropy is non-decreasing determine
the form of the stress tensor and the currents. Let us introduce
diffusion currents $q^\mu_I$,
\eq
J^\mu_I + \nabla_{\sigma}\M_I^{\sigma\mu} = r_I u^\mu+q^\mu_I\,,
\eeq
with the constraint $u_\mu q_I^\mu=0$, meaning that the diffusive process is
purely spatial according to an observer comoving with the fluid element. Next
we introduce  the heat flux $q^\mu$ and the viscous stress tensor, decomposed
in a symmetric traceless part $\tau^{\mu\nu}$ and a trace $\tau$, such that
\eq
T^{\mu\nu}=({\cal P}+\tau)\Pi^{\mu\nu}+\rho u^{\mu}u^{\nu}-\M_I^{\mu\lambda}
          {F^{I\nu}}_{\lambda}+q^{\mu}u^{\nu}+q^{\nu}u^{\mu}+\tau^{\mu\nu}\,.
\feq
Again these quantities are purely spatial, and verify the constraints
\eq
u_\mu q^\mu=0,\qquad u_\mu\tau^{\mu\nu}=0\,.
\eeq
Note that the new fields $q_I^\mu$, $q^\mu$, $\tau$ and $\tau^{\mu\nu}$ are
first order in $\nabla u$. Finally the entropy flux is a linear combination of
all the available vectors that are at most first order in $\nabla u$. Therefore
\eq
J_S^\mu=su^\mu+\beta q^\mu-\lambda^Iq_I^\mu\,,
\eeq
where the scalars $\beta$ and $\lambda^I$ are functions of the thermodynamic
parameters.

To determine the most general form allowed for the newly introduced
quantities we impose the second law of thermodynamics. To have a
non-decreasing total entropy the four-divergence of the entropy flux
must be positive, \eq \nabla_\mu J_S^\mu\geq 0\,. \eeq Projecting
the MHD equations on $u^\mu$ we obtain \eq u_\mu\nabla_\nu
T^{\mu\nu}=u^\mu J_I^\nu F^I_{\mu\nu}\,, \eeq which can be cast,
with some straightforward manipulations, in the
form\footnote{\label{foot:susc}To get this, one has to use
$\M_I^{\mu\lambda}=\chi_{IJ}F^{J\mu\lambda}$, with the
susceptibilities $\chi_{IJ}=\chi_{JI}$, as well as the Bianchi
identities for $F^I$.}
\begin{eqnarray}
(\rho+{\cal P})\vartheta &=& -\tau\vartheta-u^\mu\nabla_\mu\rho+u_\mu u^\nu
\nabla_\nu q^\mu-\nabla_\mu q^\mu+u_\mu\nabla_\nu\tau^{\mu\nu}\nonumber \\
&& -\frac12 u^{\mu}\M_I^{\nu\lambda}\nabla_{\mu}F^I_{\nu\lambda} -
u^\mu F^I_{\mu\nu}(J^{\nu}_I+\nabla_{\sigma}\M_I^{\sigma\nu})\,, \label{rho+P}
\end{eqnarray}
with the expansion $\vartheta=\nabla_\mu u^\mu$.

We have the Euler relation
\eq
\rho+{\cal P}=s\T+\mu^Ir_I\,,
\label{euler}\eeq
which implies the Gibbs-Duhem relation
\eq
\dd {\cal P} = s\,\dd\T+r_I\,\dd\mu^I + \M_I\dd {\cal B}^I\,, \label{Gibbs-Duhem}
\eeq
with the magnetization densities $\M_I$. This yields
\eq
s\nabla_\mu \T=\nabla_\mu {\cal P}-r_I\nabla_\mu\mu^I-\M_I\nabla_{\mu}{\cal B}^I\,.
\eeq
The entropy density is a function of $\rho$, $r_I$ and ${\cal B}^J$, so that its
gradient reads
\eq
\nabla_\mu s=\lp\frac{\p s}{\p \rho}\rp_{r_I,{\cal B}^J}\nabla_\mu\rho+\lp
\frac{\p s}{\p r_I}\rp_{\rho,{\cal B}^J}\nabla_\mu r_I+\lp\frac{\p s}{\p{\cal B}^J}
\rp_{r_I,\rho}\nabla_\mu{\cal B}^J=\frac1\T\nabla_\mu\rho-\frac{\mu^I}\T\nabla_\mu
r_I+\frac{\M_J}{\T}\nabla_{\mu}{\cal B}^J\,.
\eeq
Then
\begin{eqnarray}
\nabla_\mu J_S^\mu &=& \frac1\T u^\mu\nabla_\mu\rho-\frac{\mu^I}\T u^\mu
\nabla_\mu r_I+\frac{\M_J}{\T}u^{\mu}\nabla_{\mu}{\cal B}^J+\lp\frac{{\cal P}+\rho-
\mu^I r_I}{\T}\rp\vartheta \nonumber \\
&& +\beta\nabla_\mu q^\mu+q^\mu\nabla_\mu\beta-\lambda^I\nabla_{\mu}q^\mu_I-
q^\mu_I\nabla_\mu\lambda^I\,.
\end{eqnarray}
The conservation of the charge currents $\nabla_\mu J^\mu_I=0$ yields the
relation\footnote{Note that the polarization current is separately conserved,
$\nabla_{\mu}\nabla_{\sigma}\M^{\sigma\mu}=0$.}
\eq
u^\mu\nabla_\mu r_I=-\nabla_\mu q^\mu_I-r_I\vartheta\,,
\eeq
that, together with (\ref{rho+P}), can be used to put the divergence of the
entropy flux in the form
\begin{eqnarray}
&&\nabla_\mu J_S^\mu=q^\mu\lp\nabla_\mu\beta-\frac1\T u^\nu\nabla_\nu u_\mu\rp+
\lp\beta-\frac1\T\rp\nabla_\mu q^\mu-\lp\lambda^I - \frac{\mu^I}{\T}\rp
\nabla_\mu q^\mu_I\nonumber\\
&&\qquad\qquad -q^\mu_I\lp\nabla_\mu\lambda^I-\frac1\T
F_{\mu\nu}^Iu^\nu\rp-\frac\tau\T\vartheta-\frac1\T\tau^{\mu\nu}\nabla_\nu
u_{\mu}\,. \label{divJS}
\end{eqnarray}
Note that, in order to get (\ref{divJS}), we assumed that the
grandcanonical potential $\Phi$ depends on $F^I$ only through the
invariant $F^I_{\mu\nu}F^{I\mu\nu}$ (no summation over $I$), and we {\it defined}
$({\cal B}^I)^2$ to be given (up to a prefactor) by $F^I_{\mu\nu}F^{I\mu\nu}$.
This leads to
\begin{equation}
-\frac 12 \M_I^{\nu\lambda}\nabla_{\mu}F^I_{\nu\lambda} =
-\M_I\nabla_{\mu}{\cal B}^I\,.
\label{MnablaF}
\end{equation}
We ensure that the right hand side of (\ref{divJS}) is positive or vanishing
by requiring that each term be positive or vanishing. This can be easily
achieved by choosing
\eq
\beta = \frac1\T\,,\qquad
\lambda^I = \frac {\mu^I}\T\,.
\eeq
Then, the diffusion currents are given by
\eq
q^\mu_I=-D_{IJ}\Pi^{\mu\nu}\left[\nabla_\nu\lp\frac{\mu^J}\T\rp-\frac1\T
F^J_{\nu\sigma}u^\sigma\right]\,, \label{diff-curr}
\eeq
where we have introduced the projector $\Pi^{\mu\nu}$ to ensure that the
diffusion currents be spacelike, and the diffusion matrix $D_{IJ}$ is
positive definite. Moreover, we set
\eq
\tau=-\zeta\nabla_\mu u^\mu\,,
\eeq
where $\zeta\geq0$ is the bulk viscosity, and the heat flux is given by
\eq
q^\mu=-\kappa \T\Pi^{\mu\nu}\lp\frac1\T\nabla_\nu\T+u^\si\nabla_\si u_\nu\rp\,,
\label{heatflux}
\eeq
with $\kappa\geq0$ the thermal conductivity of the fluid.
Finally,
\eq
\tau^{\mu\nu}=-\eta\sigma^{\mu\nu}\,,
\eeq
with $\eta\geq0$ the shear viscosity of the fluid and $\sigma^{\mu\nu}$ the
shear tensor,
\eq
\si^{\mu\nu}=\frac12\lp\Pi^{\si\nu}\nabla_\si u^\mu+\Pi^{\si\mu}\nabla_\si
u^\nu\rp-\frac13\Pi^{\mu\nu}\vartheta\,.
\eeq
With these definitions, we find
\eq
\nabla_\mu J_S^\mu=\frac{q^\mu q_\mu}{\ka\T^2}+\frac{\zeta}{\T}\vartheta^2+\lp
D^{-1}\rp{}^{IJ}\,q^\mu_I q_{J\mu} + \frac{\tau^{\mu\nu}\tau_{\mu\nu}}{\eta\T}
\geq0\,,
\eeq
which is positive by construction. Notice finally that the stress tensor in
(\ref{T-J-JS}) is traceless on account of the equation of state
(\ref{state-equ}).

%%%%%%%%%%%%%%%%%%%%%%%%%%%%%%%%%%%%%%%%%%%%%%%%%%%%%%%%%%%%%%%%%%%%%%%%%%
\subsection{MHD equations from gravity}
\label{MHD-grav}

In this section, we consider Einstein-Maxwell-AdS gravity in $D$ dimensions and show that at leading order the boundary theory dual to a charged $D$-dimensional AdS black hole reduces to magnetohydrodynamics in $d=D-1$ dimensions in the long wavelength sector. We follow the usual procedure to foliate asymptotically the spacetime with timelike
hypersurfaces $\Sigma_r$ and regularize the action by adding appropriate boundary counter\-terms. Then, in the $r\rightarrow\infty$ limit, the Brown-York stress tensor has a finite limit, the renormalized holographic stress tensor of the
dual CFT. The projection of the Einstein-Maxwell equations on $\Sigma_r$ then
shows that this stress tensor together with the boundary gauge fields satisfy
the equations of magnetohydrodynamics on the boundary. We can interpret this as
the leading contribution in the derivative expansion to the boundary MHD
equations describing the long wavelength sector of gravity.

The Einstein-Maxwell action with negative cosmological constant
$\Lambda=-(D-2)(D-1)/2\ell^2$ reads
\eq
I=\frac1{16\pi G}\int d^Dx\sqrt{-g}\lp R-2\Lambda- \F_{MN}\F^{MN}\rp
+I_{CS} + \frac1{8\pi G}\int_{\Si_R}d^dx\sqrt{-h}\,K + I_{ct}\,,
\eeq
where $I_{CS}$ is the Chern-Simons term, present in the $D=5$ case.
$M,N,\ldots$ are bulk indices, whereas $\mu,\nu,\ldots$ refer to the boundary.
In this subsection (and only here), we refer to the bulk gauge field as $\F_{MN}$ to distinguish it from the boundary gauge field. In all other sections of the paper we do not need such
a distinction.
$\Si_R$ denotes the boundary hypersurface $r=R$ with outward pointing unit
normal $n^M$, and induced metric
\eq
h_{MN}=g_{MN}-n_Mn_N\,.
\feq
$K$ is the trace of the extrinsic curvature defined by (with $\DD_M$ the bulk covariant derivative)
\eq
K_{MN}=h_M{}^P\DD_Pn_N\,.
\eeq
Finally, $I_{ct}$ are the usual boundary counterterms needed to obtain
a finite action in the limit $R\rightarrow\infty$. We shall not need the
precise form of $I_{ct}$, but only the fact that
their variation with respect to the metric is divergence-free,
\eq
\hat \nabla^\mu\frac{\delta I_{ct}}{\delta h^{\mu\nu}}=0\,.
\label{divct}\eeq
Here we have defined $\hat \nabla_\mu$ as the induced covariant derivative on $\Si_R$.
Also, notice that additional terms can be added to $I_{ct}$ to handle the
logarithmic divergences appearing for odd $D$, corresponding to the Weyl anomaly
of the dual CFT.
We can use the Fefferman-Graham expansion to write the metric of any asymptotically AdS spacetime near spatial infinity in the form
\eq
ds^2=\frac{\ell^2}{r^2}dr^2+r^2g_{\mu\nu}(r,x)dx^\mu dx^\nu,
\eeq
where
\eq
g_{\mu\nu}=g^{(0)}_{\mu\nu}+\frac1{r^2}g^{(2)}_{\mu\nu}+\cdots+\frac1{r^d}
g^{(d)}_{\mu\nu}+h^{(d)}_{\mu\nu}\frac{\ln r}{r^d}+{\cal O}\lp\frac1{r^{d+1}}\rp\,.
\eeq
The coefficients $g^{(a)}_{\mu\nu}$ and $h^{(d)}_{\mu\nu}$ depend only on the boundary coordinates $x^\mu$ and the coefficient $h^{(d)}_{\mu\nu}$, related to the holographic Weyl anomaly, is
present for odd $D$ only.
The normal vector to the constant $r$ hypersurfaces is $n^A\p_A=r/\ell\,\p_r$ and the induced metric is $h_{\mu\nu}=\left.r^2g_{\mu\nu}\right|_{r=R}$. Then, the conformal boundary metric of AdS$_D$ is obtained by taking the limit
\eq
\ga_{\mu\nu}=\lim_{R\rightarrow\infty}\frac{\ell^{2}}{R^{2}}\,h_{\mu\nu}\,.
\eeq
Let us decompose the gauge field in the orthogonal component $\hat J^M$ and its
projection $\hat F_{MN}$ on $\Sigma_R$,
\eq
\F_{MN}=\hat F_{MN}+\frac 12(\hat J_Mn_N-\hat J_Nn_M)\,,
\eeq
where
\eq
\hat J^M=2h^{MN}n^P\F_{NP}\,,\qquad
\hat F_{MN}=h_M{}^Ph_N{}^Q\F_{PQ}\,.
\label{hats}\eeq
The analysis of the asymptotic behavior of vector gauge fields \cite{Ishibashi:2004wx,Marolf:2006nd} shows that $\hat F_{\mu\nu}$ has a finite $R\rightarrow\infty$ limit, while the current $\hat J^M$ goes to zero like $r^{-d}$. Therefore, the renormalized background gauge field in the dual field theory, and the $R$-symmetry current read respectively
\eq
F_{\mu\nu}=\lim_{R\rightarrow\infty}\hat F_{\mu\nu}\,,\qquad
J^\mu=\lim_{R\rightarrow\infty}\frac{\sqrt{-h}}{\sqrt{-\gamma}}\hat J^\mu\,.
\eeq

Now, by varying $I$ with respect to the bulk metric and gauge field, we obtain the
equations of motion\footnote{In $D=5$ one has to take into account that there is a Chern-Simons term and the Maxwell equations read $\dd^\star \F+\frac2{\sqrt3}\F\wedge \F=0$.}
\eq
E_{MN}=G_{MN}+\Lambda g_{MN}-\T_{MN}=0\,,\qquad \DD^M\F_{MN}=0\,,
\eeq
where the stress tensor of the gauge field is given by
\eq
\T_{MN}=2\F_M{}^P\F_{PN}-\frac12\F^2g_{MN}\,.
\eeq
Then, from the projection
\eq
E_{MN}n^M{h^N}_P=0
\feq
of the Einstein equations, we obtain, using Gauss-Codazzi
\eq
\hat\nabla_M\lp K^M{}_N-\delta^M{}_N K\rp=\hat F_{NM}\hat J^M\,.
\label{einstein2}\eeq
We recognize in the term in parenthesis the Brown-York boundary stress tensor, which diverges as we take the $R\rightarrow\infty$ limit, but will give a finite limit -- the holographic stress tensor $T^{\mu\nu}$ -- once we take into account the counterterm contribution \cite{Balasubramanian:1999re},
\eq
\sqrt{-\ga}\,\ga^{\mu\nu}T_{\nu\rho}=\lim_{R\rightarrow\infty}\sqrt{-h}\,h^{\mu\nu}
\lp K_{\nu\rho}-h_{\nu\rho}K+\frac{\delta I_{ct}}{\delta h^{\nu\rho}}\rp,
\eeq
Since the counterterm contribution is divergence-free by (\ref{divct}),
we can add its divergence to the left hand side of (\ref{einstein2}) and, after multiplying the equation by $\sqrt{-h}$, obtain a finite $R\rightarrow\infty$ limit that reads
\eq
\nabla_\mu T^\mu{}_\nu=F_{\nu\mu}J^{\mu}\,,
\label{mhd}\eeq
where $\nabla_\mu$ is the covariant derivative induced on the conformal boundary with metric $\ga_{\mu\nu}$.
On the other hand, using the definition (\ref{hats}) it follows that
$\hat\nabla_\mu\hat J^\mu=0$, that becomes, in the $R\rightarrow\infty$ limit, the conservation law for the $R$-current,
\eq
\nabla_\mu J^\mu=0\,.
\label{rcurr}\eeq
Note that equations (\ref{mhd}) and (\ref{rcurr}) are simply the Ward identities  associated to the bulk diffeomorphism and gauge invariance respectively \cite{Bianchi:2001kw}.

These conservation equations for the boundary stress-energy tensor and $R$-current become the equations for magnetohydrodynamics in the background field $F_{\mu\nu}$ when a large black hole is present in the bulk. Indeed, as noted in \cite{Awad:1999xx}, stationary Kerr-AdS black holes have a holographic stress tensor that assumes the perfect fluid form. We will show in the next section that this is still the case for dyonic Kerr-Newman-AdS$_4$ black holes, and we believe it to be true in any dimension. This dual stress tensor, of the perfect fluid form, yields, when combined with equations (\ref{mhd}) and (\ref{rcurr}), the equations of MHD. Then, if one perturbs these stationary solutions with long wavelength disturbances, in the spirit of \cite{Bhattacharyya:2008jc}, the horizon can still be decomposed into patches that tubewise extend to the boundary which are approximated by boosted pieces of black branes. Therefore, to leading order in the derivative expansion, 
large magnetically charged black holes are dual to a magnetohydrodynamic theory.
We shall check this explicitely in the next section for dyonic AdS$_4$ black holes. Higher orders in the perturbation theory produce higher-derivative dissipative terms in the stress tensor. A detailed analysis of this long wavelength sector of gravity, with a complete proof of the duality with MHD and the computation of the dual stress tensor up to third order in the derivative expansion has been performed by Hansen and Kraus in the AdS$_4$ case and appeared during the last stages of preparation of this manuscript \cite{Hansen:2008tq}.

%%%%%%%%%%%%%%%%%%%%%%%%%%%%%%%%%%%%%%%%%%%%%%%%%%%%%%%%%%%%%%%%%%%%%%%%%%%%%
\setcounter{equation}{0}\section{AdS black holes and black strings from MHD}

\subsection{Static dyonic black holes in AdS$_4$}

The equations of motion following from the Einstein-Maxwell action with
negative cosmological constant $\Lambda=-3\ell^{-2}$,
\begin{equation}
I = \frac 1{16\pi G}\int d^4x\sqrt{-g}\left[R - F_{MN}F^{MN} -
2\Lambda\right]\,, \label{EM-action}
\end{equation}
admit the static dyonic black hole solutions
\begin{equation}
ds^2 = -V(r)dt^2 + \frac{dr^2}{V(r)} + r^2(d\theta^2+S(\theta)^2d\phi^2)\,,
\end{equation}
where
\eq
V(r) = \frac{r^2}{\ell^2} + k - \frac{2m}r + \frac{q_e^2+q_m^2}{r^2}\,,
\feq
and
\eq
S(\theta) = \left\{\begin{array}{r@{\,,\quad}l} \sin\theta & k=1\,, \\ 1 &
k=0\,, \\ \sinh\theta & k=-1\,. \end{array} \right. \label{S(theta)}
\feq
The horizon is thus $S^2$ ($k=1$), $\bR^2$ ($k=0$) or $H^2$ ($k=-1$).
$m$, $q_e$ and $q_m$ denote the mass parameter, electric and magnetic
charge respectively. The one-form gauge potential reads
\eq
A_t = -\frac{q_e}r\,, \qquad A_{\phi} = q_m\int S(\theta) d\theta\,.
\feq
The strength of the magnetic field $B$ in the dual CFT can be obtained
(up to rescaling by powers of $\ell$) by
taking $r\to\infty$ in the expression for the bulk field strength.
This leads to $B=q_m/\ell^3$. The electric charge density
$\rho_e$ of the state in the field theory is given by
$\rho_e=\langle J^t\rangle$, where $J^{\mu}$ is the R-current that can be computed
as follows. On-shell we have for the variation of the action with respect
to the gauge potential
\eq
\frac{\delta I}{\delta A_N}\delta A_N = -\frac 1{4\pi G}\int d^4x
\partial_M(\sqrt{-g}F^{MN}\delta A_N) = -\frac 1{4\pi G}\int d^3x
\sqrt{-h}n_M F^{MN}\delta A_N\,,
\feq
where $h$ denotes the induced metric on the boundary, and $n$ is the outward
pointing unit normal to the boundary. One has thus in the limit of large $r$
\eq
\frac{\delta I}{\delta A_t} = \frac 1{4\pi G}\frac{\ell q_e}{r^3}\,.
\feq
To get the CFT R-current, one has to rescale this by $r^3/\ell^2$, so that
\eq
\langle J^t\rangle = \frac{\sqrt 2 N^{3/2}q_e}{6\pi\ell^3}\,, \label{Jt}
\feq
where we used the AdS/CFT dictionary
\eq
\frac 1{16\pi G} = \frac{\sqrt 2 N^{3/2}}{24\pi\ell^2}\,.
\feq
Note that the result (\ref{Jt}) was obtained for $k=0$
in \cite{Hartnoll:2007ai}. In order for the potential to be regular at the
horizon $r=r_h$, $A_t$ must vanish there. This requires that we add the
pure gauge term $(q_e/r_h\ell)dt$ to $A/\ell$. This term is non-normalizable
and has the dual interpretation of adding a chemical potential for the electric
charge, $\zeta=q_e/r_h\ell$, to the field theory \cite{Hartnoll:2007ai}.
The R-charge is $R = \rho_e V$, with the spatial volume\footnote{In order to
get a finite volume in the cases $k=0$ and $k=-1$, one has to compactify the
horizon to a torus or a higher genus Riemann surface respectively.}
\eq
V = \ell^2\int d\theta d\phi S(\theta)\,.
\feq
This yields
\eq
R = \frac{\sqrt 2 N^{3/2}q_e V}{6\pi\ell^3}\,. \label{R}
\feq
The entropy $S$ and energy $E$ of the black hole read
\eq
S = \frac{A_h}{4G} = \frac{\sqrt 2 N^{3/2}r_h^2V}{6\ell^4}\,, \label{S}
\feq
\eq
E = \frac m{4\pi G}\frac V{\ell^2} = \frac{\sqrt 2N^{3/2}V}{12\pi\ell^4}
\left[\frac{r_h^3}{\ell^2} + kr_h + \frac{q_e^2+q_m^2}{r_h}\right]\,.
\feq
In what follows, we shall be interested in a magnetohydrodynamical
description of the above black holes. Like in \cite{Bhattacharyya:2007vs},
we may estimate the mean free path for the fluid as
$l_{\mathrm{mfp}}\sim\eta/\rho$, where $\eta$ is the shear viscosity and
$\rho$ is the energy density of the fluid. For fluids described by a
gravitational dual, one has $\eta=s/4\pi$, where $s$ is the entropy
density \cite{Son:2007vk}. Consequently, $l_{\mathrm{mfp}}\sim s/4\pi\rho$,
and an MHD description will be valid if this value is much smaller than
the radius of the $S^2$ or the $H^2$ curvature radius\footnote{In
the (compactified) $k=0$ case, $l_{\mathrm{mfp}}$ must be much smaller than
the length of any torus cycle.}, $s/4\pi\rho \ll \ell$. Using the expressions
for $S$ and $E$, this implies
\eq
\frac{r_h}{\ell} + k\frac{\ell}{r_h} + \frac{(q_e^2+q_m^2)\ell}{r_h^3}\gg 1\,.
\label{estimate}
\feq
A sufficient condition for (\ref{estimate}) to hold is $r_h/\ell\gg 1$,
i.~e.~, for large black holes. In this case we can neglect the
$k$-contribution to the energy, and the thermodynamic fundamental relation
becomes
\eq
E(S,V,R,B) = \frac{\sqrt 2N^{3/2}V}{12\pi}\left(\frac{6S}{\sqrt 2N^{3/2}V}
\right)^{3/2}\left[1 + \left(\frac{\pi R}S\right)^2 + \frac{B^2N^3V^2}{18S^2}
\right]\,. \label{fund-rel}
\feq
One easily checks that
\eq
\frac{\partial E}{\partial S} = T\,, \qquad \frac{\partial E}{\partial R}
= \zeta\,,
\feq
where
\eq
T = \frac{V'(r_h)}{4\pi} = \frac 1{4\pi}\left[\frac{3r_h}{\ell^2} -
\frac{q_e^2+q_m^2}{r_h^3}\right] \label{T-stat}
\feq
is the Hawking temperature of the black hole. From (\ref{fund-rel}), we
get the grandcanonical potential
\eq
\Phi(T,V,\zeta,B) = E - TS - \zeta R = -VT^3h(\zeta/T,B/T^2)\,, \label{Phi3d}
\feq
with the function $h$ given by
\eq
h = \frac{\sqrt 2N^{3/2}}{24\pi}\left[H^3 + H\left(\frac{\zeta}T\right)^2
- \frac 3H\left(\frac B{T^2}\right)^2\right]\,, \label{h}
\feq
where $H(\zeta/T,B/T^2)=r_h/(\ell^2T)$ is determined by the fourth order equation
\eq
3H^4 - 4\pi H^3 - (\zeta/T)^2H^2 - (B/T^2)^2 = 0\,, \label{H}
\feq
that follows from (\ref{T-stat}).
Note that $\Phi$ has indeed the form (\ref{Phi}), as it must be.

We now want to obtain the dyonic AdS$_4$ black holes from MHD on
$\bR\times S^2$, $\bR\times\bR^2$ or $\bR\times H^2$. The metric on the
conformal boundary is given by
\eq ds^2 = -dt^2 + \ell^2(d\theta^2 + S(\theta)^2 d\phi^2)\,,
\feq
so that the only nonzero Christoffel symbols are
\eq {\Gamma^{\theta}}_{\phi\phi} =
-S(\theta)S'(\theta)\,, \qquad {\Gamma^{\phi}}_{\theta\phi} =
{\Gamma^{\phi}}_{\phi\theta} = \frac{S'(\theta)}{S(\theta)}\,.
\label{Christoffel}
\feq
For stationary, translationally invariant and axisymmetric configurations
one has $\partial_t T^{\mu\nu} = \partial_{\phi} T^{\mu\nu} = 0$ (we
also assume $\partial_t F^{\mu\nu} = \partial_{\phi} F^{\mu\nu} = 0$), and the
MHD equations $\nabla_{\mu} T^{\mu\nu} = {F^{\nu}}_{\mu}J^{\mu}$ become thus
\begin{eqnarray}
&&\partial_{\theta} T^{\theta t} + \frac{S'}S T^{\theta t} = {F^t}_{\mu}J^{\mu}\,,
\label{MHD3d-t} \\
&&\partial_{\theta} T^{\theta\theta} + \frac{S'}S T^{\theta\theta} - SS'
T^{\phi\phi} = {F^{\theta}}_{\mu}J^{\mu}\,, \label{MHD3d-theta} \\
&&\partial_{\theta} T^{\theta\phi} + \frac{3S'}S T^{\theta\phi} =
{F^{\phi}}_{\mu}J^{\mu}\,. \label{MHD3d-phi}
\end{eqnarray}
The macroscopic electric charge current and the entropy current are given
respectively by
\eq
J^{\mu}_{\mathrm{macr.}} = J^{\mu} + \nabla_{\sigma}\M^{\sigma\mu} = ru^{\mu}\,,
\qquad J^{\mu}_S = su^{\mu}\,, \label{RS-curr}
\feq
where $u^{\mu}=\gamma(1,\vec v)$ is the 3-velocity
of the fluid, $r$ denotes the electric charge density and $s$ is
the rest frame entropy density. Both currents are conserved,
\eq
\nabla_{\mu} J^{\mu}_{\mathrm{macr.}} = \nabla_{\mu} J^{\mu}_S = 0\,.
\feq
As there are no dissipative terms in the charge- and entropy currents, we
have for the entropy
\eq
S = \int d^2 x\sqrt{-g} J^t_S = \int d\theta d\phi\,\ell^2 S(\theta)s\gamma\,,
\label{S-fluid}
\feq
and for the electric charge
\eq
R = \int d^2 x\sqrt{-g} J^t_{\mathrm{macr.}} = \int d\theta d\phi\,\ell^2
    S(\theta)r\gamma\,. \label{R-fluid}
\feq
The Killing vectors of interest are $\partial_t$ (energy $E$) and
$\partial_{\phi}$ (angular momentum $J$ on the $S^2$, $\bR^2$ or $H^2$). The
conserved charge related to a Killing vector $k$ is proportional to
$\int d^2x \sqrt{-g} \,{T^t}_{\mu} k^{\mu}$, and hence
\begin{eqnarray}
E &=& \int d\theta d\phi\,\ell^2 S(\theta) T^{tt}\,, \label{E-fluid}\\
J &=& \int d\theta d\phi\,\ell^4 S^3(\theta) T^{t\phi}\,. \label{J-fluid}
\end{eqnarray}
The shear tensor $\sigma^{\mu\nu}$, heat flux $q^\mu$ and diffusion current
$q_D^\mu$ must vanish on any stationary solution of fluid dynamics. The
requirement $\sigma^{\mu\nu}=0$ means that the fluid motion should be just a
rigid rotation. By an SO$(3)$ transformation\footnote{Strictly speaking, an
applied electromagnetic field breaks SO$(3)$ invariance, so that our choice
of $u^{\mu}$ implies consistency conditions on $F_{\mu\nu}$ (cf.~(\ref{equ_T_mu})
below).}
we can can choose this rotation
such that the 3-velocity of the fluid is $u^\mu=(u^t, u^{\theta}, u^{\phi})=
\gamma(1, 0, \omega)$ for some constant $\omega$. From $u^\mu u_\mu=-1$ and
$\gamma=(1-v^2)^{-1/2}$ one obtains then $v^2=\omega^2\ell^2S^2(\theta)$.

The equilibrium fluid flow is symmetric under translations of $t$
and $\phi$, so that all thermodynamic quantities depend only on
$\theta$. We now evaluate the expansion, acceleration, shear tensor, heat flux
(\ref{heatflux}) and diffusion current (\ref{diff-curr}), with the result
\eq
\vartheta=0\,,\qquad a^{\mu}=(0, -S(\theta)S'(\theta)\gamma^2\omega^2, 0)\,,
\qquad \sigma^{\mu\nu}=0\,,
\eeq
\eq
q^\mu=-\kappa\ell^{-2}\gamma\lp 0, \frac d{d\theta}\lp\frac\T\gamma\rp, 0\rp\,,
\eeq
\eq
q^\mu_D=-D\ell^{-2}\lp -\frac 1\T F_{t\phi}\gamma\omega, \frac d{d\theta}\lp
\frac{\mu}\T\rp-\frac1\T F_{\theta\sigma}u^{\sigma}, \frac 1\T F_{t\phi}\gamma
\rp\,,
\eeq
where $\kappa$ and $D$ denote the thermal conductivity and the diffusion
coefficient respectively, ${\cal T}$ is the local temperature and $\mu$ the
local chemical potential. The requirement that $q^{\mu}$ and $q^{\mu}_D$ vanish
implies that
\eq
{\cal T}=\tau\gamma\,, \qquad F_{t\phi}=0\,, \qquad \frac d{d\theta}\lp
\frac{\mu}\ga\rp - \frac 1{\gamma}F_{\theta\sigma}u^{\sigma}=0\,, \label{equ_T_mu}
\eeq
with $\tau$ constant. If $F_{\theta\sigma}u^{\sigma}=0$, the last equ.~is solved by
$\mu = {\cal T}\nu$, where $\nu$ is constant. The conditions (\ref{equ_T_mu})
determine all the thermodynamic quantities as a function of the
coordinate $\theta$. We now want to shew that this configuration
solves the equations of magnetohydrodynamics. To this end, we first
notice that the dissipative part of the energy-momentum tensor
vanishes once (\ref{equ_T_mu}) is imposed, so that all nonzero
contributions to the stress tensor result from the perfect fluid
part plus interaction with the external electromagnetic field, and read
\begin{eqnarray}
T^{\mu\nu} = \gamma^2\left(\begin{array}{c@{\;\;}cc} \rho+v^2{\cal P} &
                        0 & (\rho+{\cal P})\omega \\
                        0 & \gamma^{-2}\ell^{-2}{\cal P} & 0 \\
                        (\rho+{\cal P})\omega & 0 & \rho\omega^2 + {\cal P}
                        \ell^{-2}S^{-2}(\theta)\end{array}\right) +
                        T^{\mu\nu}_{\mathrm{int.}}\,, \label{Tmunu-fluid}
\end{eqnarray}
where ${\cal P}$ denotes the local pressure and $T^{\mu\nu}_{\mathrm{int.}}$
is given by
\begin{eqnarray}
T^{\mu\nu}_{\mathrm{int.}} = -\M^{\mu\lambda}{F^{\nu}}_{\lambda} = \left(
\begin{array}{ccc} \M^{t\theta}F_{t\theta} & 0 & -\M^{\theta\phi}F_{t\theta} \\
0 & -\ell^{-2}(\M^{t\theta}F_{t\theta}+\M^{\theta\phi}F_{\theta\phi}) & 0 \\
-\M^{\theta\phi}F_{t\theta} & 0 & -\ell^{-2}S^{-2}(\theta)\M^{\theta\phi}
F_{\theta\phi}\end{array}\right)\,.
\end{eqnarray}
Eqns.~(\ref{MHD3d-t}) and (\ref{MHD3d-phi}) imply
${F^t}_{\mu}J^{\mu}={F^{\phi}}_{\mu}J^{\mu}=0$, which are automatically
satisfied if the susceptibility $\chi$ does not depend on $t$ and $\phi$,
which we assume in the following. The only nontrivial
equation of motion (\ref{MHD3d-theta}) becomes
\eq
\frac{d{\cal P}}{d\theta}
- ({\cal P}+\rho)\frac{d\ln\gamma}{d\theta} - \frac 12\M^{\nu\lambda}
\nabla_{\theta}F_{\nu\lambda} = rF_{\theta\mu}u^{\mu}\,. \label{MHD3dprime}
\feq
Using the Gibbs-Duhem relation (\ref{Gibbs-Duhem}) as well as (\ref{MnablaF}),
equ.~(\ref{MHD3dprime}) can be cast into the form
\eq
\ga s\frac{d}{d\th}\lp\frac\T\ga\rp +r\ga\frac
d{d\th}\lp\frac{\mu}\ga\rp = rF_{\theta\mu}u^{\mu}\,,
\eeq
which is automatically solved using (\ref{equ_T_mu}). Therefore, the
rigidly rotating configurations are stationary solutions of the
magnetohydrodynamic equations. Moreover, as the diffusion current and
the heat flux vanish, the only nonzero contributions to the transport
R-charge- and entropy currents come from the perfect fluid pieces
(\ref{RS-curr}), which are easily seen to be conserved as well.

From the grandcanonical potential (\ref{Phi}) we find
\begin{eqnarray}
{\cal P} &=& {\cal T}^3h(\nu,b)\,, \qquad {\cal M} = {\cal T}\frac{\partial h}
{\partial b}\,, \qquad s = {\cal T}^2\left(3h - \nu\frac{\partial h}
{\partial\nu} - 2b\frac{\partial h}{\partial b}\right)\,, \nonumber \\
\rho &=& 2({\cal P} - {\cal M}{\cal B}) = 2{\cal T}^3\left(h - b\frac{\partial h}
{\partial b}\right)\,, \qquad r = {\cal T}^2\frac{\partial h}{\partial\nu}\,.
\end{eqnarray}
In the nonrotating case $\omega=0$, we have $\gamma=1$,
$F_{\theta\sigma}u^{\sigma}=F_{\theta\phi}u^{\phi}=F_{\theta\phi}\gamma\omega=0$,
${\cal B}=B=\mathrm{const.}$, and thus (\ref{equ_T_mu})
implies that the local temperature $\cal T$ as well as $\nu$ and $b$ are
constant. We can then easily compute the energy, entropy, R-charge and
magnetization, with the result
\begin{eqnarray}
E &=& 2\tau^3V\left(h-b\frac{\partial h}{\partial b}\right)\,, \qquad
R = \tau^2V\frac{\partial h}{\partial\nu}\,, \nonumber \\
S &=& \tau^2V\left(3h-\nu\frac{\partial h}{\partial\nu}-2b\frac{\partial h}
{\partial b}\right)\,, \qquad M = \tau V\frac{\partial h}{\partial b}\,.
\end{eqnarray}
It is straightforward to verify\footnote{One can express
$dE-\tau dS-\tau\nu dR+MdB$ in terms of $d\tau$, $d\nu$, $db$ and check that it
vanishes.} that the Hawking temperature of the black
hole is given by $T=\tau$, and the chemical potential associated to the
R-charge is $\zeta=\tau\nu$.
Using the relations
\eq
\frac{\partial h}{\partial\nu} = \frac{\sqrt 2 N^{3/2}}{6\pi}H\nu\,, \qquad
\frac{\partial h}{\partial b} = -\frac{\sqrt 2 N^{3/2}}{6\pi}\frac bH\,,
\label{aux-rel}
\feq
following from (\ref{h}), (\ref{H}), one easily shows that $E$, $S$ and
$R$ coincide with the corresponding expressions (\ref{fund-rel}), (\ref{S})
and (\ref{R}) for the static black hole.

In the next subsection we will show that, using the function $h$ determined
from the static AdS$_4$ black hole as an input into the MHD equations, one can
exactly reproduce the thermodynamics, boundary stress tensor and R-current
of the rotating dyonic Kerr-Newman-AdS$_4$ solution.

\subsection{Kerr-Newman-AdS$_4$}

The Kerr-Newman-AdS$_4$ solution to the equations of motion following from (\ref{EM-action})
is given by\footnote{For the time being, we restrict to the case of spherical horizons. The
generalization to $k=0,-1$ is straightforward.}
\eq
ds^2 = -\frac{\Delta_r}{\rho^2}[dt - \frac{a\sin^2\theta}{\Xi}d\phi]^2 + \frac{\rho^2}{\Delta_r}dr^2
             + \frac{\rho^2}{\Delta_{\theta}}d\theta^2 + \frac{\Delta_{\theta}\sin^2\theta}{\rho^2}
              [adt - \frac{r^2+a^2}{\Xi}d\phi]^2\,, \label{KNAdS}
\feq
\eq
A = -\frac{q_er}{\rho^2}[dt - \frac{a\sin^2\theta}{\Xi}d\phi] + \frac{q_m\cos\theta}{\rho^2}
       [adt - \frac{r^2+a^2}{\Xi}d\phi]\,,
\feq
where
\eq
\Delta_r = (r^2+a^2)\left(1+\frac{r^2}{\ell^2}\right) - 2mr + q_e^2 + q_m^2\,, \qquad \Xi = 1 - \frac{a^2}{\ell^2}\,,
\feq
\eq
\rho^2 = r^2 + a^2\cos^2\theta\,, \qquad \Delta_{\theta} = 1 - \frac{a^2}{\ell^2}\cos^2\theta\,,
\feq
and $a$ is a rotation parameter.

The thermodynamical phase structure and stability of these black holes was studied in detail
in \cite{Caldarelli:1999xj}. In order to get the boundary geometry $d\sigma^2$, one takes $r=\mathrm{const.}\to\infty$,
and then rescales $ds^2$ by $\ell^2/r^2$. This leads to
\eq
d\sigma^2 = -[dt - \frac{a\sin^2\theta}{\Xi}d\phi]^2 + \frac{\ell^2}{\Delta_{\theta}}d\theta^2 +
\frac{\ell^2\Delta_{\theta}\sin^2\theta}{\Xi^2}d\phi^2\,, \label{rot-Einstein}
\feq
which is a rotating Einstein universe. In the limit of large $r$, the U(1) field
strength behaves as
\eq
F \to\frac{q_m}{\Xi}\sin\theta\,d\theta\wedge d\phi\,. \label{F-rot}
\feq
As before, the expectation value of the CFT R-current $J^{\mu}$ can
be computed by varying the action on-shell with respect to the gauge potential
$A^{\mu}$ and subsequently rescaling by $r^3/\ell^2$. This yields
\eq
J_t = -\frac{\sqrt 2 N^{3/2}q_e}{6\pi\ell^3}\,, \qquad
J_{\phi} = \frac{\sqrt 2 N^{3/2}q_e a\sin^2\theta}{6\pi\ell^3\Xi}\,, \qquad
J_{\theta} = 0\,. \label{J-KN}
\feq
The chemical potential $\zeta$ of the dual field theory is given
by \cite{Caldarelli:1999xj}
\eq
\zeta\ell = A^e_{\nu}\chi^{\nu}|_{r\to\infty} - A^e_{\nu}\chi^{\nu}|_{r=r_h}
= \frac{q_e r_h}{r_h^2 + a^2}\,, \label{mu1}
\feq
where $\chi=\partial_t+\Omega_H\partial_{\phi}$ denotes the null generator of the
horizon and $A^e_{\nu}$ is the electric part of the vector potential.
The angular velocity of the horizon reads
\eq
\Omega_H = \frac{a\Xi}{r_h^2+a^2}\,.
\feq
In what follows, we shall be interested in the limit of large black holes,
$r_h\gg\ell$, when a hydrodynamical description is valid. As $a^2<\ell^2$
(otherwise $\Delta_{\theta}$ can become negative), we have then also
$r_h^2\gg a^2$, and the chemical potential (\ref{mu1}) reduces to
\eq
\zeta = \frac{q_e}{\ell r_h}\,. \label{mu2}
\feq
The electric charge of the black hole (the field theory R-charge) is obtained
by computing the flux of the electromagnetic field strength at infinity,
with the result\footnote{Note that we rescaled $R$ by a factor of $\ell$
and $\zeta$ by $\ell^{-1}$ with respect to \cite{Caldarelli:1999xj}.}
\eq
R = \frac{\ell q_e}{\Xi G}\,. \label{R-rot}
\feq
The temperature, Bekenstein-Hawking entropy, energy and angular momentum are
given by \cite{Caldarelli:1999xj}
\begin{eqnarray}
T &=& \frac {r_h}{4\pi(r_h^2+a^2)}\left(1+\frac{a^2}{\ell^2}+\frac{3r_h^2}
      {\ell^2}-\frac{a^2+q_e^2+q_m^2}{r_h^2}\right)\approx \frac 1{4\pi r_h}
      \left(\frac{3r_h^2}{\ell^2}-\frac{q_e^2+q_m^2}{r_h^2}\right)\,,
      \label{T-rot} \\
S &=& \frac{\pi(r_h^2+a^2)}{\Xi G}\approx \frac{2\pi\sqrt 2 N^{3/2}r_h^2}
      {3\ell^2\Xi}\,, \label{S-rot} \\
E &=& \frac{m}{\Xi^2 G} \approx
      \frac{\sqrt 2 N^{3/2}}{3\ell^2\Xi^2}\left[\frac{r_h^3}{\ell^2} +
      \frac{q_e^2+q_m^2}{r_h}\right]\,, \label{E-rot} \\
J &=& aE\,, \label{J-rot}
\end{eqnarray}
where in the last steps we took the large black hole limit.
Let us write the metric (\ref{KNAdS}) in the ADM form
\eq
ds^2 = -N^2dt^2 + \frac{\rho^2}{\Delta_r}dr^2 + \frac{\rho^2}{\Delta_{\theta}}
       d\theta^2 + \frac{\Sigma^2\sin^2\theta}{\rho^2\Xi^2}
       (d\phi-\Omega dt)^2\,,
\feq
where
\eq
\Sigma^2 = (r^2+a^2)^2\Delta_{\theta} - a^2\Delta_r\sin^2\theta\,,
\feq
and the lapse function $N$ and the angular velocity $\Omega$ are defined by
\begin{eqnarray}
N^2 &=& \frac{\rho^2\Delta_r\Delta_{\theta}}{\Sigma^2}\,, \nonumber \\
\Omega &=& \frac{a\Xi}{\Sigma^2}\left[\Delta_{\theta}(r^2+a^2) - \Delta_r
           \right]\,.
\end{eqnarray}
It was shown in \cite{Caldarelli:1999xj} that the angular velocity $\omega$
entering the thermodynamics is
\eq
\omega = \Omega_H - \Omega_{\infty} = \frac{a(1+r_h^2/\ell^2)}{r_h^2+a^2}\,,
\feq
which boils down to
\eq
\omega\approx a/\ell^2\label{om-KN}
\feq
for large black holes.

By applying the implicit coordinate transformation \cite{Hawking:1998kw}\footnote{We apologize for using the same symbol
$\Phi$ for an angular coordinate and the grandcanonical potential,
but the meaning should be clear from the context.}
\eq
T=t\,, \qquad \Phi = \phi + \frac a{\ell^2}t\,, \qquad y\cos\Theta = r\cos\theta\,, \qquad
y^2 = \frac 1{\Xi}[r^2\Delta_{\theta} + a^2\sin^2\theta]
\label{coord-transf}
\feq
to the solution (\ref{KNAdS}), one gets a different slicing such that the
boundary geometry is that of a static Einstein universe,
\eq
d\tilde\sigma^2 = -dT^2 + \ell^2(d\Theta^2 + \sin^2\Theta d\Phi^2)\,.
\label{stat-Einstein}
\feq
Note that (\ref{stat-Einstein}) is related to (\ref{rot-Einstein}) by a
diffeomorphism plus a Weyl rescaling: On the boundary $r\to\infty$
(which implies $y\to\infty$), the coordinate transformation (\ref{coord-transf})
reduces to
\eq
T=t\,, \qquad \Phi = \phi + \frac a{\ell^2}t\,, \qquad \sin\Theta =
\frac{\sin\theta}{\sqrt{\Delta_{\theta}}}\,. \label{coord-transf-bdry}
\feq
Applying this to (\ref{stat-Einstein}) and subsequently rescaling with
$e^{2\varphi}=\Delta_{\theta}/\Xi$ yields (\ref{rot-Einstein}). The two
boundary metrics are thus conformally related. For the five-dimensional
Kerr-AdS black hole, this was first noticed in \cite{Bhattacharyya:2008ji}.

In the new coordinates, the Faraday tensor on the boundary becomes
\eq
F \to q_m\left(1-\frac{a^2}{\ell^2}\sin^2\Theta\right)^{-3/2}\sin\Theta\,
d\Theta\wedge\left(d\Phi - \frac a{\ell^2}dT\right)\,. \label{F}
\feq
Due to the nonvanishing component $F_{\Theta T}$, we have now also an external
electric field. This is not surprising; it results from the boost in
(\ref{coord-transf-bdry}).

In order to proceed, we need the thermodynamic variable $\cal B$. We already
stated in section \ref{stress-fluid} that ${\cal B}^2$ is proportional to
$F_{\mu\nu}F^{\mu\nu}$. To fix the prefactor, note that for the field
strength (\ref{F}), one gets
\eq
F_{\mu\nu}F^{\mu\nu} = \frac{2q_m^2}{\ell^4}\left(1-\frac{a^2}{\ell^2}\sin^2\Theta
\right)^{-2}\,, \label{F^2}
\feq
which reduces to $F^2=2q_m^2/\ell^4=2\ell^2{\cal B}^2$ in the static case $a=0$.
We have thus\footnote{Up to the prefactor $\ell$, that stems from the fact
that the boundary Faraday tensor is obtained from the corresponding bulk
quantity by projecting on the boundary and then rescaling with $\ell^{-1}$,
in flat spacetime this would just be the statement that $F^2=2({\cal B}^2-
{\cal E}^2)$, and we {\it defined} ${\cal B}^2$ to be what is usually called
${\cal B}^2-{\cal E}^2$.}
\eq
{\cal B} = \frac{\sqrt{F^2}}{\sqrt 2\ell}\,.
\feq
The grandcanonical potential $\Phi(\T,\VV,\mu,{\cal B})$ is given by
(\ref{Phi3d}), with $h(\nu,b)$ defined in (\ref{h}), (\ref{H}), and
\eq
b = \frac{\cal B}{\T^2} = \frac{\sqrt{F^2}}{\sqrt 2\ell\T^2}\,. \label{b}
\feq
One can then compute the polarization tensor $\M^{\mu\nu}$, with the result
\eq
\M^{\mu\nu} = -\frac 1{\VV}\frac{\partial\Phi}{\partial F_{\mu\nu}} =
\chi F^{\mu\nu}\,,
\feq
with the susceptibility
\eq
\chi = \frac 1{\ell^2 b\T}\frac{\partial h}{\partial b} = -\frac{\sqrt2N^{3/2}}
{6\pi H\ell^2\T}\,.
\feq
Note that $\chi$ is always negative, so that our fluid is diamagnetic.
Moreover, for large (local) temperatures $\T$ (keeping $\mu$ and $\cal B$
finite), (\ref{H}) gives $H(\mu/\T,{\cal B}/\T^2)\to4\pi/3$, and thus
\eq
\chi\to-\frac C{\T}\,, \qquad C\equiv\frac{\sqrt2N^{3/2}}{8\pi^2\ell^2}\,,
\feq
i.~e., the susceptibility obeys a diamagnetic Curie law with Curie constant $C$.

Let us now consider the MHD equations
$\nabla_{\mu}T^{\mu\nu}={F^{\nu}}_{\mu}J^{\mu}$ with the field strength
(\ref{F}). One easily shows that ${F^{\nu}}_{\mu}u^{\mu}=0$ for $\nu=T,\Phi$,
and that ${F^{\Theta}}_{\mu}u^{\mu}$ is proportional to $\omega-a/\ell^2$.
Let us assume that this vanishes, i.~e.~, that the angular velocity of the
fluid is given by $\omega=a/\ell^2$, as is suggested by (\ref{om-KN}).
Then the transport current $ru^{\mu}$ is orthogonal to the electromagnetic
field, and (\ref{equ_T_mu}) implies that $\nu=\mu/\T$ is constant.
Moreover, from (\ref{b}) and (\ref{F^2}) we have
\eq
b = \frac{q_m}{\ell^3\T^2}\left(1-\frac{a^2}{\ell^2}\sin^2\Theta\right)^{-1}
    = \frac{q_m\gamma^2}{\ell^3\T^2}\qquad\mathrm{for}\quad \omega=
    \frac a{\ell^2}\,,
\feq
which is constant by virtue of (\ref{equ_T_mu}). Using this,
it is straightforward to shew that the polarization current
$J^{\mu}_{\mathrm{micr.}}=-\nabla_{\sigma}\M^{\sigma\mu}$ is proportional to
$u^{\mu}$ in this case, and hence orthogonal to ${F^{\nu}}_{\mu}$ as well.
This means that ${F^{\nu}}_{\mu}J^{\mu}=0$; due to orthogonal magnetic and
electric fields compensating each other, there is no net Lorentz force
acting on the fluid, and the MHD equations boil down to
$\nabla_{\mu}T^{\mu\nu}=0$.
It is interesting to consider this situation from the point of view of a
reference frame which is moving with the charged fluid (the rotating Einstein
universe (\ref{rot-Einstein})). In this reference frame
the fluid is static and therefore not subject to the magnetic force. Since
the net electromagnetic force, which in this frame consists only of the
electric force, must be zero, the electric field must vanish in the moving
reference frame. This is indeed the case, as can be seen from (\ref{F-rot}).

It would be very interesting to see to which
kind of black holes the fluid configurations with $\omega$ different from
$a/\ell^2$ correspond to. They might have to do with a nontrivial nut
parameter, but we shall leave this point for future investigations.

One can now proceed to compute the conserved charges. From (\ref{S-fluid})
- (\ref{J-fluid}) one obtains
\eq
S = \frac{4\pi\ell^2\tau^2}{\Xi}\left(3h-\nu\frac{\partial h}{\partial\nu}
    -2b\frac{\partial h}{\partial b}\right)\,, \qquad
R = \frac{4\pi\ell^2\tau^2}{\Xi}\frac{\partial h}{\partial\nu}\,,
\feq
\eq
E = \frac{8\pi\ell^2\tau^3}{\Xi^2}\left(h-b\frac{\partial h}{\partial b}
    \right)\,, \qquad
J = \frac{8\pi\omega\ell^4\tau^3}{\Xi^2}\left(h-b\frac{\partial h}{\partial b}
    \right)\,,
\feq
together with the magnetization
\eq
M = \int d^2x\sqrt{-g}\M\gamma^{-1} = 4\pi\ell^2\tau
    \frac{\partial h}{\partial b}\,.
\feq
The temperature $T$ and chemical potential $\zeta$ are given by $T=\tau$ and
$\zeta=\tau\nu$ respectively, while the intensive variable conjugate to $M$
is $B=\tau^2b/\Xi$. To show this, one can express
$dE - \tau dS - \tau\nu dR + M dB - \omega dJ$ in terms of $d\tau$, $d\nu$,
$db$, $d\omega$ and check that it vanishes.
Note that $T$, $\zeta$ and $B$ are distinct from $\T$, $\mu$ and $\cal B$.
While the former quantities are asssociated to the whole fluid
configuration, the latter are local thermodynamic properties of the fluid
that vary on the boundary manifold. For the grandcanonical partition function
one finds
\eq
\ln{\cal Z}_{gc} = -\frac 1T(E-TS-\omega J-\zeta R) = \frac{VT^2h(\zeta/T,
B\Xi/T^2)}{\Xi}\,,
\feq
which means that $\ln{\cal Z}_{gc}$ of the rotating fluid is obtained simply
by dividing the grandcanonical partition function of the static fluid
by the universal factor $\Xi=1-\omega^2\ell^2$, and replacing $B$ by $B\Xi$.
This generalizes the results of \cite{Bhattacharyya:2007vs} to nonvanishing
magnetic fields.

We must now compare the fluid charges with the corresponding black hole
quantities. To do this, we first note that for large KNAdS black holes,
the temperature has the same dependence on $r_h$ as in the static case
(compare (\ref{T-rot}) with (\ref{T-stat})),
so that the function $H=r_h/(\ell^2T)$ is again determined by (\ref{H}).
Using this together with (\ref{h}) and (\ref{aux-rel}), one easily shows that
the fluid charges $R$, $S$, $E$, $J$ exactly coincide with the expressions
for the black hole given in (\ref{R-rot})-(\ref{J-rot}).
This coincidence extends also to the energy-momentum tensor and the
R-current: The holographic stress tensor of the KNAdS black hole was
determined in \cite{Caldarelli:1999xj}, and reads
\eq
T_{tt} = \frac{2m}{8\pi G\ell^2}\,, \qquad T_{t\phi} = -\frac{2ma\sin^2\theta}
{8\pi G\Xi\ell^2}\,,
\feq
\eq
T_{\phi\phi} = \frac{m\sin^2\theta[\ell^2+3a^2\sin^2\theta-a^2]}
{8\pi G\Xi^2\ell^2}\,, \qquad T_{\theta\theta} = \frac m{8\pi G\Delta_{\theta}}\,,
\feq
and all other components vanishing.
This corresponds to the boundary geometry (\ref{rot-Einstein}), i.~e.~, to the
rotating Einstein universe. To get the stress tensor for the static
Einstein universe (\ref{stat-Einstein}), recall that the latter is related to
the former by a diffeomorphism plus a Weyl rescaling,
$d\sigma^2=e^{2\varphi}d\tilde\sigma^2$, with $e^{2\varphi}=\Delta_{\theta}/\Xi$.
Under a Weyl rescaling, the energy-momentum tensor transforms as
$T_{\mu\nu}=e^{-(d-2)\varphi}\tilde T_{\mu\nu}$ (see
e.~g.~\cite{Bhattacharyya:2007vs}), which yields (taking $d=3$)
\eq
\tilde T_{TT} = \frac{m\gamma^3}{8\pi G\ell^2}(3\gamma^2-1)\,, \qquad
\tilde T_{T\Phi} = -\frac{3ma\sin^2\Theta}{8\pi G\ell^2}\gamma^5\,,
\feq
\eq
\tilde T_{\Phi\Phi} = \frac{m\sin^2\Theta}{8\pi G}\gamma^3(3\gamma^2-2)\,,
\qquad \tilde T_{\Theta\Theta} = \frac{m\gamma^3}{8\pi G}\,,
\feq
which is easily shown to exactly coincide with the fluid stress tensor
(\ref{Tmunu-fluid}).

The R-current (\ref{J-KN}) transforms as $\tilde J^{\mu}=e^{-d\varphi}J^{\mu}$
under a Weyl rescaling \cite{Bhattacharyya:2007vs}. This gives for the
R-current corresponding to the static boundary geometry
\eq
\tilde J^T = \gamma^3\frac{\sqrt 2N^{3/2}q_e}{6\pi\ell^3}\,, \qquad
\tilde J^{\Phi} = \frac a{\ell^2}\tilde J^T\,, \qquad \tilde J^{\Theta} = 0\,.
\feq
This is again equal to the transport current $J^{\mu}_{\mathrm{macr.}}=ru^{\mu}$
of the fluid.

%%%%%%%%%%%%%%%%%%
\subsection{Black strings in AdS$_5$}

We now solve the Navier-Stokes equations on
$\bR\times S^1\times S^2$, $\bR\times S^1\times\bR^2$ or $\bR\times
S^1\times H^2$. This will yield predictions for black strings in AdS$_5$,
for which at present only partial results are known \cite{Chamseddine:1999xk,
Klemm:2000nj,Copsey:2006br,Mann:2006yi,Brihaye:2007vm,Bernamonti:2007bu,
Delsate:2008iv}.
Since the function $h(\nu,b)$ entering (\ref{Phi}) is unknown for $d=4$,
$b\neq 0$, we shall consider hydrodynamics rather than MHD, i.~e.~, we will
restrict to the case of zero electromagnetic fields in this section.

The metric entering the MHD equations is
\eq
ds^2 = -dt^2 + dz^2 + \ell^2(d\theta^2 + S(\theta)^2 d\phi^2)\,,
\feq
where $S(\theta)$ was given in (\ref{S(theta)}).
The only nonzero Christoffel symbols are the ones in (\ref{Christoffel}).
For stationary, translationally invariant and
axisymmetric configurations one has $\partial_t T^{\mu\nu} =
\partial_z T^{\mu\nu} = \partial_{\phi} T^{\mu\nu} = 0$, and the Navier-Stokes
equations reduce to (\ref{MHD3d-t})-(\ref{MHD3d-phi}) (with $F_{\mu\nu}=0$)
plus the additional condition
\eq
\partial_{\theta} T^{\theta z} + \frac{S'}S T^{\theta z} = 0\,.
\feq
The entropy and the R-charges read
\eq
S = \int d^3 x\sqrt{-g} J^t_S = \int dz d\theta d\phi\,\ell^2S(\theta)s\gamma\,,
\feq
\eq
R_I = \int d^3 x\sqrt{-g} J^t_I = \int dz d\theta d\phi\,\ell^2S(\theta)r_I
      \gamma\,.
\feq
Now the Killing vectors of interest are $\partial_t$ (energy $E$),
$\partial_{\phi}$ (angular momentum $J$ on the $S^2$, $\bR^2$ or $H^2$), and
$\partial_z$ (momentum $p$ along the string). The associated conserved charges
are
\begin{eqnarray}
E &=& \int dz d\theta d\phi\,\ell^2S(\theta) T^{tt}\,, \\
J &=& \int dz d\theta d\phi\,\ell^4S^3(\theta) T^{t\phi}\,, \\
p &=& \int dz d\theta d\phi\,\ell^2S(\theta)T^{tz}\,.
\end{eqnarray}
The shear tensor $\sigma^{\mu\nu}$, heat flux $q^\mu$ and diffusion currents
$q^\mu_I$ must vanish on any stationary solution of fluid dynamics. The
requirement $\sigma^{\mu\nu}=0$ means that the fluid motion should be just a
rigid rotation. By an SO$(3)$ transformation we can choose this rotation such
that the 4-velocity of the fluid is $u^\mu=(u^t, u^z, u^{\theta}, u^{\phi})=
\gamma(1, \omega_1, 0, \omega_2)$ for some constants $\omega_1, \omega_2$. From
$u^\mu u_\mu=-1$ and $\gamma=(1-v^2)^{-1/2}$ one obtains then $v^2=\omega_1^2+
\ell^2\omega_2^2S^2(\theta)$.

The equilibrium fluid flow is symmetric under translations of $t$,
$z$ and $\phi$, so that all thermodynamic quantities depend only on
$\theta$. We now evaluate the expansion, acceleration, shear tensor,
heat flux and diffusion current, with the result
\eq
\vartheta=0\,,\qquad a^\mu=(0, 0,
-S(\theta)S'(\theta)\gamma^2\omega_2^2, 0)\,,\qquad \sigma^{\mu\nu}=0\,,
\eeq \eq q^\mu=-\ell^{-2}\kappa\gamma\lp0, 0, \frac
d{d\theta}\lp\frac\T\gamma\rp, 0\rp\,, \eeq \eq q^\mu_I=
-\ell^{-2}D_{IJ}\lp0, 0, \frac d{d\theta}\lp\frac{\mu^J}\T\rp, 0\rp\,.
\eeq
The requirement that $q^{\mu}$ and $q^{\mu}_I$ vanish implies that
\eq
{\cal T}=\tau\gamma\,, \qquad \mu^I = \T\nu^I\,, \label{taunu}
\eeq
with $\tau$ and $\nu^I$ constant. The conditions (\ref{taunu})
determine all the thermodynamic quantities as a function of the
coordinate $\theta$. We now want to shew that this configuration
solves the Navier-Stokes equations. To this end, we proceed as in
\cite{Bhattacharyya:2007vs}, and first
notice that the dissipative part of the energy-momentum tensor
vanishes once (\ref{taunu}) is imposed, so that all nonzero
contributions to the stress tensor result from the perfect fluid
part, and read
\begin{eqnarray}
\lefteqn{T^{\mu\nu} =}\nonumber \\
                   &&\gamma^2\left(\begin{array}{c@{\;\;}ccc} \rho+v^2{\cal P} & (\rho + {\cal P})\omega_1 & 0 &
                        (\rho+{\cal P})\omega_2 \\
                        (\rho+{\cal P})\omega_1 & \rho\omega_1^2+{\cal P}(1-g^{-2}\omega_2^2S^2(\theta)) & 0 &
                        (\rho+{\cal P})\omega_1\omega_2 \\
                        0 & 0 & \gamma^{-2}g^2{\cal P} & 0 \\
                        (\rho+{\cal P})\omega_2 & (\rho+{\cal P})\omega_1\omega_2 & 0 & \rho\omega_2^2 +
                        {\cal P}g^2S^{-2}(\theta)(1-\omega_1^2) \end{array}\right)\,. \nonumber
\end{eqnarray}
The only nontrivial equation of motion (\ref{MHD3d-theta}) becomes
\eq
\frac{d{\cal P}}{d\theta}
- ({\cal P}+\rho)\frac{d\ln\gamma}{d\theta} = 0\,. \label{MHDprime}
\feq
Using the Gibbs-Duhem relation (\ref{Gibbs-Duhem}), (\ref{MHDprime}) can be
cast into the form
\eq
\ga s\frac{d}{d\th}\lp\frac\T\ga\rp +r_I\ga\frac
d{d\th}\lp\frac{\mu^I}\ga\rp = 0\,,
\eeq
which is automatically solved using (\ref{taunu}). Therefore, the
rigidly rotating configurations are stationary solutions of the
equations of fluid dynamics.

From the grandcanonical potential
\eq
\Phi = -{\cal V}\T^4 h(\nu^I)
\feq
one obtains
\eq
\rho=3\calP=3\T^4h(\nu^I)\,,\qquad r_I=\T^3h_I(\nu^J)\,,\qquad
s=\T^3(4h-\nu^Ih_I)\,,
\eeq
with
\eq
h_I=\frac{\p h}{\p\nu^I}\,.
\eeq
We saw in the previous section that for large AdS$_4$ black holes, the
function $h$ is insensitive to the curvature parameter $k$. The same is to be
expected for large black strings in AdS$_5$, so that we can infer the
function $h(\nu^I)$ from the $k=0$ case, which is simply the AdS$_5$ black
hole with flat horizon. This yields \cite{Bhattacharyya:2007vs}
\eq
h(\nu^I) = 2\pi^2N^2\frac{\prod_J(1+\kappa^J)^3}{(2+\sum_J\kappa^J-\prod_J
           \kappa^J)^4}\,,
\feq
where the auxiliary parameters $\kappa^I$ are related to the $\nu^I$ by
\eq
\nu^I = \frac{2\pi\prod_J(1+\kappa^J)}{2+\sum_J\kappa^J-\prod_J\kappa^J}
        \frac{\sqrt{\kappa^I}}{1+\kappa^I}\,.
\feq

The conserved charges corresponding to our fluid configurations are
\eq
E=L\ell^2\tau^4h(\nu^I)
\lp4K_{1,6}(\om_1,\om_2)-K_{1,4}(\om_1,\om_2)\rp\,, \eeq \eq
p=4L\ell^2\tau^4\om_1h(\nu^I) K_{1,6}(\om_1,\om_2)\,,\qquad
J=4L\ell^2\tau^4\om_2h(\nu^I) K_{3,6}(\om_1,\om_2)\,, \eeq \eq
S=L\ell^2\tau^3\lp4h-\nu^Ih_I\rp K_{1,4}(\om_1,\om_2)\,,\qquad
R_I=L\ell^2\tau^3h_I(\nu^J) K_{1,4}(\om_1,\om_2)\,,
\eeq
where $L$ is the length of the $S^1$ parametrized by $z$, and we have defined
the integrals
\eq
K_{m,n}(\om_1,\om_2)=\int S^m(\theta)\gamma^n d\phi d\theta\,.
\eeq
These integrals, which can be performed in terms of elementary functions,
satisfy the relations
\eq
\frac{\p K_{m,n}}{\p\omega_1} = n\omega_1 K_{m,n+2}\,, \qquad
\frac{\p K_{m,n}}{\p\omega_2} = n\ell^2\omega_2 K_{m+2,n+2}\,, \label{der-K}
\feq
and
\eq
-\ell^2\omega_2^2 K_{m,n} = K_{m-2,n-2} - (1-\omega_1^2)K_{m-2,n}\,.
\label{rec-K}
\feq
The chemical potentials corresponding to the rotating/boosted fluid solutions
are defined by
\begin{displaymath}
T = \lp\frac{\p E}{\p S}\rp_{J,p,R_I}\,, \qquad w = \lp\frac{\p E}
{\p p}\rp_{S,J,R_I}\,, \qquad \Omega = \lp\frac{\p E}{\p J}\rp_{S,p,R_I}\,, \qquad
\zeta^I = \lp\frac{\p E}{\p R_I}\rp_{S,J,p,R_K}\,.
\end{displaymath}
One easily verifies that\footnote{As before, one can express
$dE-\tau dS-\tau\nu^IdR_I-\omega_1dp-\omega_2dJ$ in terms of
$d\tau$, $d\nu^I$, $d\omega_1$, $d\omega_2$ and check that it vanishes. In
order to show this, one has to use the relations (\ref{der-K}) and
(\ref{rec-K}).}
\eq
T=\tau\,, \qquad w=\omega_1\,, \qquad \Omega=\omega_2\,, \qquad \zeta^I =
\tau\mu^I\,.
\feq
Using this, we find for the grandcanonical partition function
\eq
\ln {\cal Z}_{gc}=-\frac1T\lp E-TS-wp-\Omega J-\zeta^IR_I\rp
=L\ell^2\tau^3h(\nu^I)K_{1,4}(\om_1,\om_2)\,.
\eeq

In the $k=0$ case, the integration is trivial and ${\cal Z}_{gc}$ reads
\eq
\ln {\cal Z}_{gc}=\frac{V\tau^3h(\nu^I)}{\lp1-\om_1^2-\ell^2\om_2^2\rp^2}\,,
\eeq
where $V$ is the volume of the three-torus.

For $k=1$, we have
\eq
\ln {\cal Z}_{gc}=\frac12V\tau^3h(\nu^I)\left[\frac1{\lp1-\om_1^2\rp\lp1-
\om_1^2-\ell^2\om_2^2\rp}+\frac{\arctan\lp\frac{\ell\om_2}{\sqrt{1-\om_1^2-
\ell^2\om_2^2}}\rp}{\ell\om_2\lp1-\om_1^2-\ell^2\om_2^2\rp^{3/2}}
\right]\,,
\eeq
where $V=4\pi L\ell^2$ is the volume of the $S^1\times S^2$. Note that this
reduces, for $\om_1=0$, to
\eq
\ln {\cal Z}_{gc}=\frac12V\tau^3h(\nu^I)\left[\frac1{1-\ell^2\om_2^2}
+\frac{\arctan\lp\frac{\ell\om_2}{\sqrt{1-\ell^2\om_2^2}}\rp}{\ell\om_2\lp1-
\ell^2\om_2^2\rp^{3/2}}\right]\,,
\eeq
while for $\om_2=0$ it becomes
\eq
\ln {\cal Z}_{gc}=\frac12V\tau^3h(\nu^I)\left[\frac1{\lp1-\om_1^2\rp^2}
+1\right]\,.
\eeq
We see that the formula correctly reproduces the $1/(1-\om^2)$ behaviour
characteristic of the rotation on spheres and the $1/(1-\om^2)^2$ behaviour
of the boost along a circle (compare also with the results of
\cite{Bhattacharyya:2007vs} for Kerr-AdS black holes and with
\cite{Hawking:1998kw}).

Our results of this section yield predictions for rotating black strings in
AdS$_5$ with momentum along the string, that carry three electric charges
(more precisely, there are electric charge densities along the string).
While the general solutions of this type are still to be discovered,
there are some partial numerical results available, namely for static charged
black strings with all three charges equal \cite{Brihaye:2007vm}. It would be
very interesting to compare the thermodynamics of these solutions with our
hydrodynamical predictions. Unfortunately, the gravity solutions have been
constructed only numerically, so that the comparison involves a fair amount
of numerical work that goes beyond the scope of this paper.

%%%%%%%%%%%%%%%%%%%%%%%%%%%%%%%%%%%%%%%%%%%%%%%%%%%%%%%%%%%%%%%%%%%%%%%%%%%%%
\setcounter{equation}{0}\section{Gregory-Laflamme on magnetized
plasma tubes \label{sec:RPmag}}
 %%%%%%%%%%%%%%%%%%%%%%%%%%%%%%%%%%%%%%
%%%%%%%%%%%%%%%%%%%%%%%%%%%%%%%%%%%%%%%%%%%%%%%%%%%%%%%%%%%%%%%%%%%%%%%%%%%%%
%\setcounter{equation}{0}\section{Gregory-Laflamme and Rayleigh-Plateau in magnetic backgrounds \label{sec:RPmag}}
 %%%%%%%%%%%%%%%%%%%%%%%%%%%%%%%%%%%%%%

%%%%%%%%%%%%%%%%%%%%%%%%%%%%%%%%%%%%%%%%%%%%%%%%%%%%%%%%%%%%%%%%%%%%%%%%%%%%%
\subsection{Description of the problem. Scherk-Schwarz compactification of a CFT  \label{sec:SScft}}
 %%%%%%%%%%%%%%%%%%%%%%%%%%%%%%%%%%%%%%

It was observed in
\cite{Aharony:2005bm,Lahiri:2007ae,Bhardwaj:2008if} that, in the
long wavelength regime, $d$-dimensional fluid dynamics is also an
effective theory describing the Scherk-Schwarz (SS) compactification
of a $(d+1)$-dimensional CFT. This theory is dual to SS
compactification of  AdS$_{d+2}$ gravity. At finite temperature the
theory has two vacuum solutions, namely i) the black brane solution
that describes the deconfined phase of the dual gauge theory and ii)
the AdS soliton that describes the confined phase
\cite{Aharony:2005bm}. In the neighborhood of the critical phase
transition temperature the two phases can cohabit in equilibrium
with a domain wall interface. Black holes in SS compactified
AdS$_{d+2}$ are then dual to deconfined plasma configurations
immersed in the confined phase. The phase diagram of plasma
equilibrium solutions includes plasma balls, pinched plasma balls,
plasma rings and plasma tubes
\cite{Lahiri:2007ae,Caldarelli:2008mv,Bhardwaj:2008if}. These plasma
solutions are dual to SS compactified AdS black objects whose
explicit solution is not yet known. However, quite interestingly,
these phase diagrams reveal to have properties that are quite
similar (although with some differences linked to the AdS nature of
the system) to the known phase diagram solutions of asymptotically
{\it flat} black holes. In \cite{Caldarelli:2008mv} we have argued
that this is not a coincidence since asymptotically flat black holes
may indeed admit a fluid description in the limit of large number of
dimensions. It is therefore important to explore the properties of
plasma objects since they might provide a lighthouse to understand
better even asymptotically flat black holes. We will return to this
discussion in the end of this section.

In a previous paper \cite{Caldarelli:2008mv}, we have shown that
plasma tubes suffer from the long wavelength Rayleigh-Plateau
instability that makes them pinch-off when their length is bigger
than their radius. We have further observed that these solutions are
the natural duals of SS AdS black strings and that the plasma
instability is dual to the Gregory-Laflamme instability of a black
string. The aforementioned analysis was constructed on the top of a
series of studies
\cite{Cardoso:2006ks,Cardoso:2006sj,Cardoso:2007ka,Dias:2007hg,Miyamoto:2008rd,Miyamoto:2008uf}
where the remarkable {\it analogy} between fluid tubes and black
strings was observed and analyzed. After
\cite{Lahiri:2007ae,Bhardwaj:2008if}, \cite{Caldarelli:2008mv}
promoted this curious analogy to an actual duality.

In this section we want to go a step forward and discuss the
effect that a magnetic background introduces in the Rayleigh-Plateau
instability of a plasma tube. This will provide solid predictions
for the features of the dual magnetic black strings, associated
Gregory-Laflamme instability and phase diagram of associated
magnetically charged solutions.

\vskip 0.2cm

We will take our boundary geometry to be $\bR_t\times
 \mathbb{R}^2\times S^1$, described by the metric
 \eq
  ds^2 = -dt^2 + dR^2+R^2d\phi^2+dz^2\,, \label{BdryGeom:RP}
\feq
 In this background, we consider fluid configurations that are
uniform plasma tubes with topology $S^1\times D^2$, \ie a disk
extended along the periodic $z$-direction and immersed in the vacuum
or confined phase. We also take them to be translationally invariant
along the SS  direction (not represented in (\ref{BdryGeom:RP})).
Furthermore, we assume that there is a constant magnetic field along
the $z$ direction. The dynamics of this plasma configuration is
governed by the equations of relativistic magnetohydrodynamics (MHD)
in $(3+1)$-dimensions. As mentioned above, in the long wavelength
regime, MHD is an effective theory describing the SS
compactification of a $(4+1)$-dimensional CFT. Moreover, in the dual
gravity description these plasma tube configurations correspond to
black strings in a background that asymptotes to a SS
compactification of AdS$_6$. The horizon topology of these strings
is $S^1\times S^3$. The reason being that on the interface between
the plasma phase and the vacuum, the SS  circle shrinks to zero size
as we enter in the holographic direction and reach the horizon
radius.

%%%%%%%%%%%%%%%%%%%%%%%%%%%%%%%%%%%%%%%%%%%%%%%%%%%%%%%%%%%%%%%%%%%%%%%%%%%%%
\subsection{MHD for Scherk-Schwarz plasmas \label{sec:MHD:SS}}
 %%%%%%%%%%%%%%%%%%%%%%%%%%%%%%%%%%%%%%

In this subsection we find the MHD equations that describe the long
wavelength magnetohydrodynamic limit of a Scherk-Schwarz
compactification of a $(d+1)$-dimensional CFT. In the end we
specialize to the $d=4$ case.

Neglecting subleading dissipation and diffusion contributions (in
particular, in equilibrium these contributions vanish) the
energy-momentum tensor of the fluid in the presence of an external
electromagnetic field is the sum of the perfect fluid, Maxwell
interaction and boundary surface contributions,\footnote{The perfect
fluid and Maxwell interaction terms were already introduced in
(\ref{T-J-JS}). In section \ref{sec:MHD}, where we studied the MHD
equations of a CFT on $\mathbb{M}^d$, these were the only
non-dissipative contributions. Here we want to study them in the
case where we have a SS compactification of a $(4+1)$-dimensional
CFT. Then there is also a surface contribution. It arises
because in this case we have an interface between the plasma and the
vacuum phases.}
\begin{eqnarray}
&&T^{\mu\nu}=T^{\mu\nu}_{\rm perf}+T^{\mu\nu}_{\rm int}+T^{\mu\nu}_{\rm bdry}\,;\nonumber\\
&&\qquad T^{\mu\nu}_{\rm perf}=\lp\rho+\calP\rp u^\mu u^\nu+\calP g^{\mu\nu}\,,\nonumber\\
&& \qquad T^{\mu\nu}_{\rm int}=-\M^{\mu\lambda}
          {F^{\nu}}_{\lambda}\,,\nonumber\\
&& \qquad T^{\mu\nu}_{\rm bdry}= -\sigma h^{\mu\nu}|\partial
f|\,\delta(f)\,.\label{GenTuv}
\end{eqnarray}
Here, $u^{\mu}$ is the fluid velocity, $\rho$, $\calP$ and $\sigma$
are the density, pressure and surface tension of the fluid,
$F_{\mu\nu}$ is the external Maxwell field and $\M^{\mu\nu}$ is
the associated polarization tensor. The fluid boundary is defined by
$f(x^{\mu})=0$, it has unit normal $n_{\mu}=\partial_{\mu}
f/|\partial f|$, and $h^{\mu\nu}=g^{\mu\nu}-n^\mu n^\nu$ is the
projector onto the boundary. % Finally, $\Theta(-f)$ is the Heaviside function ($\Theta(-f)=1$ inside the fluid and zero everywhere else).

The MHD equations follow from the covariant divergence of the stress
tensor,
\begin{equation}
\nabla_\nu T^{\mu\nu}= J_\nu F^{\mu\nu}\,,\label{DivTuv}
\end{equation}
where $J_\nu$ is the charge current defined in (\ref{T-J-JS}). From
these equations we can derive the relativistic continuity,
Navier-Stokes and Young-Laplace equations. In the absence of an
electromagnetic field this derivation was presented in detail in
\cite{Caldarelli:2008mv}. With the extra Maxwell contribution the
derivation follows similarly and we get (using footnote
\ref{foot:susc})
\begin{eqnarray}
&& \hspace{-1.5cm}u^{\mu}\partial_{\mu}\rho + (\rho+\calP)
\nabla_{\mu} u^\mu +\frac12
u^{\mu}\M^{\alpha\beta}\nabla_{\mu}F_{\alpha\beta}= 0\,,
\label{continuity} \\
&& \hspace{-1.5cm}(\rho+\calP) u^\nu \nabla_{\nu} u^\mu =
-\Pi^{\mu\nu}\lp \nabla_{\nu} \calP -\frac12
\M^{\alpha\beta}\nabla_{\nu}F_{\alpha\beta}\rp
+r u_\nu F^{\mu\nu}\,, \label{NavierS} \\
&& \hspace{-1.5cm}
 \Pi_<-\Pi_>=\sigma K\,, \qquad \Pi\equiv
 \calP-\M^{\mu\alpha}F_{\nu\alpha}n_\mu n^\nu \,, \qquad K\equiv h_{\mu}^{\:\:\nu}\nabla_{\nu}
 n^{\mu}\,,
 \label{YoungLap}            % $K=h_{\mu}^{\:\:\nu}\nabla_{\nu} n^{\mu}$
 \end{eqnarray}
where $K$ is the boundary's extrinsic curvature and $\Pi$ is the
generalized fluid pressure. It is the sum of the usual
pressure $\calP$ and the magnetic pressure (that is due to the
Lorentz force exerted by the electromagnetic field on the
polarization current). $\Pi_<-\Pi_>$ is the generalized pressure
jump when we cross the boundary from the exterior (with pressure
$\Pi_>$) into the interior (with pressure $\Pi_<$). In the
derivation of (\ref{YoungLap}), the constraint that the fluid
velocity must be orthogonal to the boundary normal is used (this
guarantees that the fluid is confined inside the boundary),
\begin{equation}
 u^\mu n_{\mu} = 0 \,. \label{constraintYL}
\end{equation}

Equations (\ref{continuity})-(\ref{constraintYL}) constitute the set
of relativistic MHD equations governing the fluid dynamics. A
particular solution of these equations describes a static uniform
plasma tube in the background (\ref{BdryGeom:RP}) with radius $R_0$ and
extended along the $z$-direction, and with a constant magnetic field
${\cal B}_{(0)}$ in the same direction. This solution is
characterized by
\begin{equation}
 u^\mu_{(0)}=\delta^{\mu}_{\:t}\,,\qquad {\cal B}_{(0)}^{\mu}={\cal B}_{(0)}\delta^{\mu}_{\:z}\,,
 \qquad \calP=\calP_{(0)}\,,\qquad \rho=\rho_{(0)}\,.\label{UNperturbed}
 \end{equation}
So, the non-vanishing Maxwell tensor components are
$F_{R\phi}=-F_{\phi R}={\cal B}_{(0)}\sqrt{-g}$ and
$F_{\mu\nu}F^{\mu\nu}=2 {\cal B}_{(0)}^2$. We use the subscript
$(0)$ to emphasize that this is a static unperturbed solution of
MHD. In these conditions some terms in
(\ref{continuity})-(\ref{YoungLap}) simplify considerably. In
particular,
$\M^{\alpha\beta}\nabla_{\nu}F_{\alpha\beta}=2\M_{(0)}\nabla_{\nu}{\cal
B}_{(0)}=0$, $r u_\nu F^{\mu\nu}=0$ and $\Pi=
 \calP-\M_{(0)}{\cal B}_{(0)}$ (where $\M_{(0)}=\chi {\cal B}_{(0)}$).
  $\calP_{(0)}$ is then a constant by the Navier-Stokes equations (\ref{NavierS}).
The pressure $\calP_{(0)}$, energy density $\rho_{(0)}$ and magnetic
field ${\cal B}_{(0)}$ are not independent. They are related through
the appropriate equation of state for a SS compactification of a
$(4+1)$-dimensional CFT. We derive this equation in the next
subsection. It is given by (\ref{SS:state-dimRed}) with $d=4$. This
equation of state together with the continuity equation
(\ref{continuity}) demands that $\rho_{(0)}$ is also a constant.

In subsection \ref{sec:RPinst} we will perturb equations
(\ref{continuity})-(\ref{constraintYL}) and study the
Rayleigh-Plateau instability of a relativistic plasma tube in the
presence of a magnetic field.

%%%%%%%%%%%%%%%%%%%%%%%%%%%%%%%%%%%%%%%%%%%%%%%%%%%%%%%%%%%%%%%%%%%%%%%%%%%%%
\subsection{Equation of state for a Scherk-Schwarz plasma \label{sec:SS:eqState}}
 %%%%%%%%%%%%%%%%%%%%%%%%%%%%%%%%%%%%%%

In this subsection we discuss the grandcanonical potential and
equation of state of a Scherk-Schwarz compactification of a
$(d+1)$-dimensional CFT. The $d=4$ result will be needed in the next
subsection.

Start with a $(d+1)$-dimensional CFT on $\mathbb{M}^d\times
\calS^1$, where $\calS^1$ is the SS  circle, in the presence of a
conserved R-charge density $r$ (and associated chemical potential
$\mu$) and of a constant magnetic field ${\cal B}$. Conformal
invariance and extensivity demand that the grandcanonical potential,
\begin{equation}
\Phi({\cal T},{\cal V}, \mu, {\cal B}) = {\cal E} - {\cal T}{\cal S}
- \mu{\cal R} \,, \label{grandcan-pot:bSS}
\end{equation}
is given by
\begin{equation}
\Phi = -{\cal V}{\cal T}^{d+1} h(\nu, b)\,, \label{SS:Phi}
\end{equation}
where we defined $\nu=\mu/{\cal T}$ and $b={\cal B}/{\cal T}^2$. So,
a CFT on $\mathbb{M}^d\times \calS^1$ has the same thermodynamical
potential as a CFT on $\mathbb{M}^{d+1}$ (that we analyzed in
section \ref{eqState}\footnote{Note that in section \ref{eqState} we
worked in $d$-dimensions while in this section we work in $d+1$
dimensions.}). The compact SS  direction does however introduce a
Casimir or vacuum energy density $\rho_0$\footnote{In what follows,
we shall assume that $\rho_0$ is independent of $\nu$ and $b$.}.
Therefore, the equation of state of the CFT on $\mathbb{M}^d\times
\calS^1$ is
\begin{equation}
\rho = d\,{\cal P} - 2{\cal B}{\cal M}+\rho_0\,.
\label{SS:state-equ}
\end{equation}
It follows from $\Phi=\Phi({\cal T},{\cal V}, \mu, {\cal B})$ and
(\ref{SS:Phi}) that the pressure of the conformal fluid on
$\mathbb{M}^{d+1}$ is given by
\begin{equation}
{\cal P}= -\frac{\partial\Phi}{\partial{\cal V}}{\biggr|}_{{\cal
T},\mu,{\cal B}}={\cal T}^{d+1} h(\nu, b)\,.
\end{equation}
Replacing this in the equation of state (\ref{SS:state-equ}) we
obtain for the local temperature
\begin{equation}
{\cal T}= \lp\frac{\rho-\rho_0+2{\cal B}{\cal M}}{d\, h(\nu,
b)}\rp^{\frac{1}{d+1}}\label{SS:temp}\,.
\end{equation}

We can now do a dimensional reduction along the SS direction. This
yields a field theory (that is not conformal) and which has a
grandcanonical potential given by
\begin{eqnarray}
\Phi_{\mathrm{red}}({\cal T},{\cal V}, \mu, {\cal B}) &=& {\cal E} - {\cal
T}{\cal S}
- \mu{\cal R}\nonumber\\
 &=& {\cal V}\lp \rho_0-{\cal T}^{d+1} h(\nu, b)\rp\,. \label{SS:grandcan-pot}
\end{eqnarray}
Upon this reduction, the local temperature of the plasma does not
change and is still given by (\ref{SS:temp}). From
$\Phi_{\mathrm{red}}=\Phi_{\mathrm{red}}({\cal T},{\cal V}, \mu, {\cal B})$ one has,
\begin{eqnarray}
&& s=-\frac{1}{\cal V}\frac{\partial\Phi_{\mathrm{red}}}{\partial{\cal
T}}={\cal T}^{d}\left[ (d+1)h(\nu, b)-\nu\partial_\nu h(\nu,
b)-2b\partial_b
h(\nu, b)\right] \,, \nonumber\\
 && {\cal P}=-\frac{\partial\Phi_{\mathrm{red}}}{\partial{\cal V}}={\cal T}^{d+1} h(\nu, b)-\rho_0 \,,
 \nonumber\\
 && r=-\frac{1}{\cal V}\frac{\partial\Phi_{\mathrm{red}}}{\partial{\cal \mu}}={\cal T}^{d} \partial_\nu h(\nu, b) \,, \nonumber\\
 && {\cal M}=-\frac{1}{\cal V}\frac{\partial\Phi_{\mathrm{red}}}{\partial {\cal B}}={\cal T}^{d-1} \partial_b h(\nu, b)
 \,.\label{SS:Thermo}
\end{eqnarray}
Replacing (\ref{SS:temp}) in the expression (\ref{SS:Thermo}) for
the pressure we get the equation of state,
\begin{equation}
{\cal P} =\frac{\rho-(d+1)\rho_0+2{\cal B}{\cal M}}{d} \,,
\label{SS:state-dimRed}
\end{equation}
which is valid in or out of equilibrium. A similar replacement could
be done for the other equations in (\ref{SS:Thermo}).

It is important to observe that the hydrodynamical equilibrium of
the magnetized uniform tube also implies its thermodynamic
equilibrium \cite{Caldarelli:2008mv}. Indeed, the static condition
$\partial_\mu\calP=0$ that follows from the Navier-Stokes equation,
together with the Gibbs-Duhem relation (\ref{Gibbs-Duhem}), written
as $\nabla_\nu {\cal P} = s\,\nabla_\nu\T+r\,\nabla_\nu \mu + \M
\nabla_\nu {\cal B}$, implies that the local temperature $\T$ is
constant. Our uniform tube is static and thus the plasma temperature
$T$ equals the local temperature $\T$. Demanding then that
(\ref{SS:temp}) is constant we find the relation (for constant $c$)
\begin{equation}
 \rho-\rho_0=c\,d\,h(\nu,b)-2{\cal B}\M
\label{SS:rhoDif}\,,
\end{equation}
for an equilibrium static solution. The two last relations will be
used later in the $d=4$ case.

%%%%%%%%%%%%%%%%%%%%%%%%%%%%%%%%%%%%%%%%%%%%%%%%%%%%%%%%%%%%%%%%%%%%%%%%%%%%%
\subsection{Rayleigh-Plateau instability in a magnetic
background\label{sec:RPinst}}
 %%%%%%%%%%%%%%%%%%%%%%%%%%%%%%%%%%%%%%

We now address the stability of a uniform plasma tube with a
constant magnetic field along the tube direction when we perturb it.
The dynamics of the perturbations are dictated by the MHD equations
subject to appropriate boundary conditions.
 Once perturbations settle in, the static plasma tube is taken away from thermal
equilibrium and viscosity and diffusion effects start being also
active. Therefore, the energy-momentum tensor of the fluid includes
now not only the perfect fluid, the Maxwell interaction and the
boundary surface tension terms (\ref{GenTuv}), but also a
dissipative contribution. The uniform plasma tube is afflicted by
the Rayleigh-Plateau instability \cite{Caldarelli:2008mv}. Surface
tension is the mechanism responsible for this instability. Viscosity
and diffusion play  no role on the activation of the instability.
Therefore, in our analysis we will neglect the dissipation
contribution to the fluid stress tensor, and comment at the end on the
effects it introduces. Our main aim is to find the effect that a
magnetic field introduces on the instability, extending the analysis
of \cite{Caldarelli:2008mv}.

Take a static uniform plasma tube with radius $R_0$ in a constant
magnetic field ${\cal B}_{(0)}$ along the tube $z$-direction. This
unperturbed solution is described by (\ref{UNperturbed}) and by the
equation of state (\ref{SS:state-dimRed}), with $d=4$.

Consider now perturbations on this plasma tube. A generic
perturbation is described as
\begin{equation}
 u^\mu=u^\mu_{(0)}+\delta u^\mu\,,\qquad \calP=\calP_{(0)}+\delta \calP\,,
 \qquad \rho=\rho_{(0)}+\delta \rho\,,\qquad {\cal B}^{\mu}={\cal B}_{(0)}^{\mu}\,,\label{perturbation}
 \end{equation}
where the perturbation on a quantity $Q$ is generically denoted by
$\delta Q$. Note that the magnetic field is kept fixed. As there are
no Maxwell equations on the boundary, it is treated as a
non-dynamical parameter. Some of these perturbations are not
independent. They are related by the equation of state
(\ref{SS:state-dimRed}), valid also out of equilibrium. Its
perturbation yields the relation between the density and pressure
perturbations,
\begin{equation}
 \delta \rho\simeq 4\delta \calP \,. \label{rhoPpert}
 \end{equation}
In deriving this, we made the assumption that
$\delta({\cal B}{\cal M})={\cal B}^2\delta\chi$ vanishes. This seems reasonable
for small magnetic fields, or in a regime where the susceptibility varies
only slowly as a function of the temperature and the other thermodynamic
variables. We shall comment later on possible effects that a relaxation
of this assumption might have.

The perturbed state must satisfy the continuity and Navier-Stokes
equations, (\ref{continuity}) and (\ref{NavierS}). The Young-Laplace
equation (\ref{YoungLap}) and the constraint (\ref{constraintYL})
provide boundary conditions for the perturbed problem. After
eliminating the $0^{\rm th}$ order terms using the unperturbed MHD
equations, the continuity and Navier-Stokes eqns.~yield, up to first order
in the perturbation,
\begin{eqnarray}
&& \hspace{-1.7cm}\partial_R \delta u^R+\frac{1}{R}\,\delta u^R+
\partial_\phi \delta u^\phi +\partial_t  \delta u^t+
\frac{1}{\rho_{(0)}+\calP_{(0)}}\,\partial_t\delta\rho  =0 \,,
\label{Pert:continuity} \\
 && \hspace{-1.7cm}\lp\rho_{(0)}+\calP_{(0)}\rp
\partial_t\delta u^\mu=-\delta^\mu_{~t}\partial_t \delta
\calP-g^{\mu\nu}\partial_\nu \delta \calP +r \delta u_R F^{\mu
R}+r\delta u_\phi F^{\mu \phi}=0\,. \label{Pert:NavierS}
\end{eqnarray}

Since any perturbation can be written as a Fourier series, in the
most general case we will consider a perturbation that disturbs the
boundary surface of the plasma tube according to
\begin{equation}
 R=R(t,z,\phi)\,, \qquad R(t,z,\phi)=R_0+ \epsilon \,e^{\omega
t}e^{ikz+im \phi} \,,\qquad \epsilon \ll R_0 \,, \label{Pert:radius}
\end{equation}
Positive $\omega$ describes an instability with wavenumber $k$. The
possible unstable mode is axisymmetric if $m=0$. Naturally, we look
for perturbations of the fluid quantities that have the same form as
the boundary disturbance,
\begin{equation}
\delta Q(t,R,z,\phi)=\delta Q(R) e^{\omega t}e^{ikz+im \phi}
\,,\qquad \delta Q\equiv\{\delta u^\mu,\delta \calP,\delta \rho \}
\,, \label{Pert:Q}
\end{equation}
which determines $\partial_{\mu}\delta Q$ for $\mu=t,z,\phi$.

Solving (\ref{Pert:NavierS}) with perturbations (\ref{Pert:Q}) and
(\ref{rhoPpert}) we find that the velocity perturbations at leading
order are
\begin{eqnarray}
&&\delta u^t (R)=0 \nonumber\\
 &&\delta u^R (R)=-
\left[\omega\lp\rho_{(0)}+\calP_{(0)}\rp \lp
1+\frac{k^2_A}{k^2}\rp\right]^{-1} \lp \frac{d\, \delta
\calP(R)}{dR}+
\frac{i m k_A}{k R}\,\delta\calP \rp \,,\nonumber\\
&& \delta u^\phi (R)=- \left[\omega\lp\rho_{(0)}+\calP_{(0)}\rp
\lp1+\frac{k^2_A}{k^2}\rp\right]^{-1} \lp -\frac{k_A}{k
R}\,\frac{d\, \delta \calP(R)}{dR}+
\frac{i m }{ R^2}\,\delta\calP(R) \rp \,, \nonumber\\
&& \delta u^z
(R)=-i\,\frac{k}{\omega\lp\rho_{(0)}+\calP_{(0)}\rp}\,\delta
\calP(R)
 \,,
\label{Pert:Q:static}
\end{eqnarray}
where we defined the Alfv\'en wavenumber of the system,
\begin{equation}
k_A\equiv \frac{{\cal B}_{(0)}
\,r}{\rho_{(0)}+\calP_{(0)}}\frac{k}{\omega}\,. \label{Pert:Alfven}
\end{equation}
Note that (\ref{Pert:Q:static}) satisfies the perturbed version of
$u^\mu u_\mu=-1$, namely $u_\mu^{(0)}\delta u^\mu=0$.

 Replacing (\ref{Pert:Q:static}) in the continuity equation
(\ref{Pert:continuity}) and using (\ref{rhoPpert}) we get
\begin{equation}
\frac{d^2 \delta \calP(R)}{dR^2}+\frac{1}{R}\frac{d\, \delta
\calP(R)}{dR}-\left[\eta^2+\frac{m^2}{R^2} \right]\delta
\calP(R)=0\,,\qquad \eta^2\equiv k^2\lp 1+\frac{4\omega^2}{k^2}\rp
 \lp 1+\frac{k^2_A}{k^2}\rp . \label{Pert:Bes:st}
\end{equation}
This is a modified Bessel equation. Its solutions are the modified
Bessel functions of the first kind, $I_{m}(\eta R)$, and second kind
$K_{m}(\eta R)$. $K_{m}(\eta R)$ diverges as $R^{-m}$ as
$R\rightarrow 0$ and we discard it (for $m=0$, this solution would
give a regular $\delta \calP$ but $\delta u^\mu$ would still
diverge; so we drop it also in this case). Therefore, the regular
solution of (\ref{Pert:Bes:st}) at the origin is
\begin{equation}
\delta \calP(R)=A I_{m}(\eta R)\,,
 \label{Pert:BesSol:st}
\end{equation}
where $A$ is a constant to be fixed. Replacing the pressure
perturbation in (\ref{Pert:Q:static}) yields for the radial
component of the velocity,
\begin{equation}
\delta u^R (R)=-\frac{A\eta}{\omega\lp\rho_{(0)}+\calP_{(0)}\rp \lp
1+\frac{k^2_A}{k^2}\rp}\,
 \lp I'_{m}(\eta R) +\frac{im}{\eta R}\,\frac{k_A}{k} \,I_m(\eta R)\rp\,,
 \label{Pert:velocitypert}
\end{equation}
where $I'_{\nu}(y)\equiv \partial_y I_{\nu}(y)$.

The solutions just found must satisfy a total of two appropriate
boundary conditions. The first  demands normal stress balance on the
boundary. This means that the generalized pressure perturbation must
also satisfy the third perturbed hydrodynamic equation, namely the
one that follows from perturbing the Young-Laplace ewquation
(\ref{YoungLap}),
\begin{equation}
{\rm BC}\:\: {\rm I:}\qquad
 \delta\Pi{\bigl |}_{\rm bdry} \simeq
 \sigma \left[ K{\bigl |}_{R(t,\phi,z)} -K{\bigl |}_{R_0} \right] \,.
\label{Pert:bc1}
\end{equation}
Here, $\delta\Pi{\bigl |}_{\rm bdry} \simeq \delta \calP{\bigl
|}_{\rm bdry}$ up to order $\epsilon^2$. The reason being that the
external magnetic field is non-dynamical and thus fixed and the
perturbation on the boundary normal is of order
$\epsilon^2$.\footnote{\label{MagYLpert}This point deserves a closer
look. One has  $\delta\Pi |_{\rm bdry} = \delta \calP |_{\rm bdry}-
\delta\lp \M^{\mu\alpha}F^{\nu}_{~~\alpha} n_\mu n_\nu\rp |_{\rm
bdry}$; see (\ref{YoungLap}). Assuming $\delta\chi \sim 0$,
consistent with the approximation taken in (\ref{rhoPpert}), the
magnetic contribution can be written as
$\M^{\mu\alpha}F^{\nu}_{~~\alpha}\delta\lp n_\mu n_\nu\rp=
\M^{\mu\alpha}F^{\nu}_{~~\alpha} ( n_{\mu}^{(0)} \delta n_\nu+
\delta n_\mu n_{\nu}^{(0)} )$, where $\delta
n_\mu=n_{\mu}|_{R(t,\phi,z)}-n_{\mu}^{(0)}$ and $n_{\mu}^{(0)}=n_\mu
|_{R_0}=\delta_{\mu}^R$. We then have $\delta\lp
\M^{\mu\alpha}F^{\nu}_{~~\alpha} n_\mu n_\nu\rp= 2\M_{(0)}{\cal
B}_{(0)}\delta n_r$. Now, using (\ref{Pert:radius}) we find that
$\delta
n_r=[(k^2-\omega^2)/2+m^2/(2R_0^2)]\epsilon^2+\mathcal{O}(\epsilon^3)
=\mathcal{O}(\epsilon^2)$. So the magnetic contribution to
$\delta\Pi |_{\rm bdry}$ is indeed of order $\epsilon^2$ and thus
subleading when compared with the rhs of (\ref{Pert:bc1}), that is
of order $\epsilon$.}
On the rhs we evaluate the expression at the
perturbed boundary $R=R(t,\phi,z)$ defined in (\ref{Pert:radius})
and we subtract the unperturbed contribution evaluated at $R=R_0$.
The extrinsic curvature  is obtained from its definition, $K=
h_{\mu}^{\:\:\nu}\nabla_{\nu} n^{\mu}$, with the unit normal of the
boundary (\ref{Pert:radius}), $n_{\mu}=|n|^{-1}\lp
-R_t'\delta_{\mu}^{\:t}+
\delta_{\mu}^{\:R}-R_{\phi}'\delta_{\mu}^{\:\phi}-R_z'\delta_{\mu}^{\:z}
\rp$. This condition fixes the ratio $A/\epsilon$ to leading order
as
\begin{equation}
\frac{A}{\epsilon}\simeq \frac{\sigma}{R_0^2 \,I_{m}(\eta R_0)}\lp
k^2 R_0^2+\omega^2R_0^2-1+m^2 \rp  \,.
 \label{Pert:Ae}
\end{equation}
The second boundary condition is a kinematic condition requiring
that the normal component of the fluid velocity on the boundary
satisfies the perturbed version of (\ref{constraintYL}),
 $u^{\mu}_{(0)}\,\delta n_\mu +\delta u^\mu n_{\mu}^{(0)}=0$, where
 $\delta n_\mu\equiv n_\mu{\bigl |}_{R(t,z,\phi)}-n_{\mu}^{(0)}$ and
the unperturbed normal is $n_\mu^{(0)}\equiv n_\mu{\bigl
|}_{R_0}=\delta_\mu^R$. This ensures that the velocity perturbation
leaves the fluid confined inside the boundary. This boundary
condition then reads
\begin{equation}
{\rm BC}\:\: {\rm II:}\qquad \delta u^R{\bigl |}_{\rm bdry}\simeq
\omega \,\epsilon \,e^{\omega t}e^{im\phi+ikz}\,, \label{Pert:bc2}
\end{equation}
where the lhs follows from (\ref{Pert:velocitypert}) evaluated at
the boundary. This boundary condition together with (\ref{Pert:Ae})
yields the desired dispersion relation $\omega (k)$ for the
Rayleigh-Plateau instability in the presence of a constant magnetic
field along the plasma tube direction,
\begin{equation}
\omega^2= \frac{\sigma}{\lp \rho_{(0)}+\calP_{(0)}\rp R_0^3}
\left[\frac{k R_0\, I_{m+1}(\eta R_0)}{I_m(\eta R_0)}
+m\lp1+i\frac{k_A}{\eta}\rp\right] \lp
1+\frac{k_A^2}{k^2}\rp^{-\frac{1}{2}}\lp 1-m^2-\omega^2R_0^2 - k^2
R_0^2\rp \,,
 \label{Pert:Dispersion}
\end{equation}
where we used the relation $I'_{\nu}(y)=I_{\nu
+1}(y)+\frac{\nu}{y}I_{\nu}(y)$ and $\eta$ is defined in
(\ref{Pert:Bes:st}). $\rho_{(0)}+\calP_{(0)}$ is obtained from the
equation of state (\ref{SS:state-dimRed}) and (\ref{SS:rhoDif}) for
$d=4$.

\begin{figure}[th]
\centerline{\includegraphics[width=.48\textwidth]{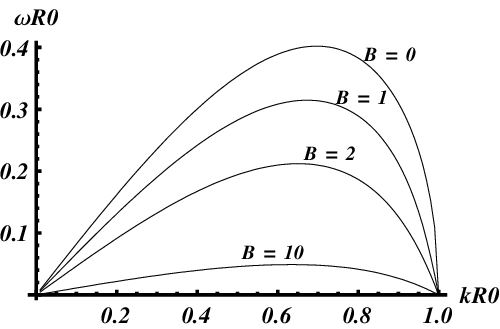}}
\caption{\small Plot of the dimensionless dispersion relation
$\omega(k)$ for the Rayleigh-Plateau instability in a static uniform
tube for several values of the axial magnetic field ${\cal B}_{(0)}$
(in the plot, $B$). The instability strength and decreases as
the magnetic field grows.
 We use (\ref{Pert:Dispersion}) and the numerical data correspond to
take $\frac{\sigma}{(\rho_{(0)}+\calP_{(0)})R_0}=\frac{10^{-6}}
 {5-2{\cal B}_{(0)}^2}$ (see text).
} \label{fig:GLmagnetic}
\end{figure}

We conclude that the Rayleigh-Plateau instability is active
($\omega>0$) for wavenumbers $k$ satisfying the condition
\begin{equation}
k\leqslant \frac{\sqrt{1-m^2}}{R_0}\,.\label{threshold}
 \end{equation}
Since $m$ is an integer this means that only axisymmetric modes
($m=0$) are unstable. In a previous paper \cite{Caldarelli:2008mv}
we have shown that the condition (\ref{threshold}) for the threshold
wavenumber can be found by noting that unstable modes are those that
reduce the potential energy of the system for fixed volume
(equivalently, they are those that maximize the entropy for fixed
conserved charges). In the present system we have a competition
between the surface tension and magnetic potential energies.
However, to leading order $\mathcal{O}(\epsilon)$  in the
perturbation, the magnetic energy does not change. Indeed, the
magnetic field is kept fixed without being dynamically perturbed,
see (\ref{perturbation}), and the volume change that also
contributes to the magnetic energy variation is subleading, \ie
$\mathcal{O}(\epsilon^2)$  (see footnote \ref{MagYLpert}), when
compared with the $\mathcal{O}(\epsilon)$ change in the surface
tension energy. Therefore, only the surface tension potential energy
can decrease (to leading order in the perturbation), and thus the
condition (\ref{threshold}) for the instability is independent of
the introduction of a background magnetic field \footnote{Later, in
the end of next subsection, we will argue that we do no expect this
to be true when the approximation $\delta\chi\sim 0$ is relaxed.}.
With or without a non-dynamical magnetic field, the Rayleigh-Plateau
instability is a long wavelength instability that afflicts the
plasma tube when its length is bigger than its transverse radius.

The presence of the non-dynamical magnetic field does however reduce
the strength of the instability. Indeed, in the plot of the
dispersion relation (\ref{Pert:Dispersion}), see
Fig.~\ref{fig:GLmagnetic}, we find that an increment in the magnetic
field ${\cal B}_{(0)}$ weakens the instability. For high values of
${\cal B}_{(0)}$ the instability strength becomes considerably weaker.
Strictly speaking we cannot however say that it disappears for a
critical value of ${\cal B}_{(0)}$ since the threshold mode
(\ref{threshold}) is always non-vanishing independently of ${\cal
B}_{(0)}$.

A word of caution is in order. From the equation of
state (\ref{SS:state-dimRed}) and the equilibrium condition
(\ref{SS:rhoDif}) one gets for $d=4$: $\rho_{(0)}+\calP_{(0)}=5a
h(\nu,b)-2{\cal B}_{(0)}\M_{(0)}$ (for constant $a$). Note that
unfortunately we do not know the function $h(\nu,b)$ in five
dimensions and this prevents an accurate computation. Depending on
its form, the condition $\rho_{(0)}+\calP_{(0)}>0$ might require a
maximum allowed ${\cal B}_{(0)}$. To plot Fig.~\ref{fig:GLmagnetic}
we took $5a h(\nu,b)\equiv 1$, \ie $\rho_{(0)}+\calP_{(0)}=5-2{\cal
B}_{(0)}\M_{(0)}$, and this indeed implies a critical ${\cal
B}_{(0)}$. However, we should have in mind that $h$ is not a
constant but $h=h(\nu,b)$. There might thus not exist a maximum
${\cal B}_{(0)}$.

%%%%%%%%%%%%%%%%%%%%%%%%%%%%%%%%%%%%%%%%%%%%%%%%%%%%%%%%%%%%%%%%
\subsection{Unstable plasma tubes and their gravitational duals
\label{RPGL}}

The magnetized plasma tubes we have been discussing are dual to
magnetic black strings in SS compactified AdS$_6$. The Rayleigh-Plateau (RP)
instablity of the plasma tube is dual to the Gregory-Laflamme (GL)
instability of the black strings. Unfortunately the magnetic black string
solutions in SS compactified AdS$_6$ have not been constructed yet, and
thus we  cannot compare our plasma results with their
gravitational duals. Therefore, our study provides solid predictions
for the features and stability of SS AdS$_6$ magnetic black strings.

There is however growing evidence that justifies the comparison of
our plasma results with the properties of magnetically charged black
strings in an asymptotically {\it flat} background. Let us pause
here to review the evidence that seems to permit this extrapolation.
In \cite{Lahiri:2007ae,Bhardwaj:2008if} the phase diagram of
axisymmetric rigidly rotating plasma configurations was found. The
set of solutions is given by plasma balls, pinched balls and plasma
rings. Quite remarkably, the entropy {\it vs} angular momentum (for
fixed energy) phase diagram displays remarkable similarities with
the phase diagram of asymptotically flat
\cite{Emparan:2001wn,Emparan:2007wm} and AdS
\cite{Caldarelli:2008pz} black hole solutions.

More closely connected with our present study,
\cite{Caldarelli:2008mv} analyzed the RP instability of a neutral
plasma tube and the associated phase diagram of non-uniform plasma
tubes and localized plasma balls. The plasma instability has
properties remarkably similar to the GL instability of a neutral,
asymptotically flat black string \cite{Gregory:1993vy} (see
\cite{Kol:2004ww,Harmark:2007md} for reviews). A brief account of
these similarities includes \cite{Caldarelli:2008mv}: (i) the
dispersion relation of the two instabilities are quite similar and,
not less important, both evolve similarly as we add extra dimensions
(compare Fig. 5 of \cite{Caldarelli:2008mv} with Fig. 3 of
\cite{Gregory:1994bj}); ii) for static configurations, only
axisymmetric modes are unstable and there is even a quantitative
match for the threshold wavenumber as the number of dimensions gets
large; for rotating solutions, non-axisymmetric modes can become
unstable in both instabilities; iii) there are similar critical
dimensions for the entropically favored solutions; iv) the phase
diagram containing the solutions living in a compact dimension,
namely (non-)uniform extended configurations and localized
configurations is astonishingly similar (compare the $S(J)$ diagram
of Fig. 1 of \cite{Caldarelli:2008mv} with Fig. 3 of
\cite{Harmark:2005pp}); the known initial time evolution of the GL
instability of an asymptotically flat black string is similar to the
dynamical evolution of the RP instability (compare Fig. 4 of
\cite{Choptuik:2003qd} with Fig. 1 of \cite{Stone}). General
arguments suggest that in the limit of large number of dimensions
the fluid description of asymptotically flat black holes should
become more and more accurate.

The just mentioned facts motivate us to compare our results for the
RP instability of a magnetized plasma tube with the know studies
\cite{Gregory:1994bj,Gregory:1994tw,Hirayama:2002hn} of the GL
instability of magnetically charged asymptotically flat black
strings/branes. In \cite{Gregory:1994tw,Hirayama:2002hn} it is found that a
magnetic charge weakens the strength of the GL instability.
In view of the above
evidence, this is in agreement with our plasma result: a magnetic
field also decreases the strength of the RP instability of the
plasma tube.

Besides showing that the GL instability gets weaker as magnetic
charge is added to the system,
\cite{Gregory:1994bj,Gregory:1994tw,Hirayama:2002hn} have also been
able to show that it actually disappears in the extremal case, when
the charge has its maximum allowed value (for some systems it is
absent even before the extreme state is reached
\cite{Hirayama:2002hn}). This property leads to an interesting new
relation. Indeed, a dynamical instability like GL is expected to be
accompanied by a local thermodynamic instability. This motivated
Gubser and Mitra \cite{Gubser:2000ec} to propose the conjecture that
a black brane/string with a non-compact translational symmetry is
classically stable if and only if it is locally thermodynamically
stable. A detailed analysis of this conjecture was further done in
\cite{Reall:2001ag}. In these studies it was indeed found that there
is a change in the sign of the specific heat of the black brane
precisely for the critical charge where the GL instability
disappears. It is thus interesting to discuss a natural version of
the Gubser-Mitra conjecture in the uniform plasma tube setup.
Unfortunately, we are not able to address this interesting issue.
The reason is two-folded. First, to compute the specific heat of the
uniform plasma tube we would need to have the relation between the
entropy and the energy of the system. To get the entropy we would
need to integrate the entropy density given by the first relation in
(\ref{SS:Thermo}). This requires knowing the precise form of the
thermodynamic function $h(\nu,b)$, defined in
(\ref{grandcan-pot:bSS}), in five dimensions. Unfortunately, this is
currently unknown. Second, although the RP instability gets weaker
as the magnetic field increases we do not find a finite critical
value above which it is not active (see however the discussion at
the end of subsection \ref{sec:RPinst}). We do find that the
instability strength reduces substantially as $B$ grows not too
large. Although in practice we can say that it disappears for large
magnetic field, strictly speaking we are not allowed to say that it
is gone since the threshold wavenumber keeps finite (see
Fig.~\ref{fig:GLmagnetic}). The RP modes become sort of marginally
stable in the sense that unstable modes are present but with very
small strength. So, at most we would be able (if we knew $h(\nu,b)$)
to check that the specific heat also never becomes positive for the
uniform plasma tube as the magnetic field grows. In the uncharged
case, $h(\nu,b)$ is a constant, and we found in
\cite{Caldarelli:2008mv} that the heat capacity is indeed always
negative. Adding a small magnetic field should not change this, but
we cannot explore the situation for larger fields.

The absence of a critical magnetic field in the RP instability
deserves however a much more careful look. As described above this
occurs because the threshold wavenumber, where the RP is marginally
stable, persists invariant under changes of the magnetic field. This
is an odd feature\footnote{Indeed, if we consider the standard RP
instability on a charged fluid tube coupled to a {\it dynamical}
magnetic field, we find that both the instability strength and the
critical wavenumber decrease when the magnetic field increases
\cite{Chandra}.} and it is not hard to convince ourselves that this
is a consequence of neglecting the susceptibility perturbations (see
discussion associated with (\ref{rhoPpert}) and footnote
\ref{MagYLpert}). Indeed, from footnote \ref{MagYLpert} we see that
when $\delta\chi\neq 0$ there is a contribution proportional to the
square of the magnetic field strength in the lhs of
(\ref{Pert:bc1}). This contribution persists in the following
equations and the threshold wavenumber (\ref{threshold}) gets then
reduced when the magnetic field increases. Unfortunately, when we
consider $\delta\chi\neq 0$, the perturbed system is left
undetermined even after using the Bianchi identities and the
conservation of the polarization current (in footnote 2). We would
need to know some microscopic information to express $\delta \chi$
in terms of perturbations of other thermodynamic quantities -- a
relation that we do not have. For this reason, and because
$\delta\chi$ is nevertheless expected to be small (as justified
previously), we took $\delta\chi\sim 0$ in our analysis. This
approximation nevertheless allowed to find that a magnetic field
decreases the RP instability strength, and we have a good control on
the effects that a non-vanishing susceptibility perturbation
introduces in the system: the instability strength still decreases
when $B$ grows and, on the top of this, the critical wavenumber also
decreases.

 \vspace{0.2cm}

Usually an unstable mode signals a bifurcation to a new branch of
configurations in the phase diagram of solutions. Without magnetic
fields this is indeed the case and the RP unstable mode of the
uniform plasma tube leads to a branch of non-uniform plasma tubes
that joins a third branch of plasma balls localized on the compact
tube direction \cite{Caldarelli:2008mv,Maeda:2008kj}. Although not
studied here, again because we do not know the function $h(\nu,b)$,
we naturally expect that similar new branches of solutions,
representing magnetic non-uniform tubes and magnetic localized
plasma balls, exist also in the phase diagram of magnetized plasma
configurations.

 \vspace{0.2cm}

To end this section, we address the effects that viscosity
introduces in the Rayleigh-Plateau instability. These were already
discussed in more detail in \cite{Caldarelli:2008mv}, so we will be
brief in our comments. Generically, at least classicaly, the
viscosity increases the wavelength of the most unstable mode, and
weakens the strength of the instability \cite{Chandra}. It has a
subleading effect on the activation of the instability (however, the
instability can get considerably weaker if the viscosity is very
high), but plays an important role at later stages in the time
evolution of the instability, namely in the pinch-off phase of the
plasma tube \cite{Stone}. Typically, the lower the viscosity is, the
higher is the number of plasma balls formed. Your fluid is
non-conformal so we have bulk and shear viscosity and thermal
dissipation. Dissipation increases considerably the technical
challenge of solving (even numerically) the MHD perturbed equations:
we get a coupled system of differential equations with a fourth
order derivative term.

%%%%%%%%%%%%%%%%%%%%%%%%%%%%%%%%%%%%%%%%%%%%%%%%%%%%%%%%%%%%%%%%5
\subsection{Regime of validity
\label{Validity}}

To conclude this section, the regime of validity of our results
should be kept in mind.

First of all, the fluid description of the deconfined plasma must be
accurate, and hence the thermodynamic quantities must vary slowly
over the mean free path $\ell_{\rm mfp}$ of the constituent
particles, which is of the order of the mass gap of the theory, or
equivalently of the order of the deconfinement temperature $T_c\sim
\frac{\rho_0}{\sigma}$. That is, all length scales $\lambda$ in the
fluid must be
\eq \lambda \gg T_c^{-1}\sim \frac{\sigma}{\rho_0}\,. \eeq
This condition imposes restrictions on the validity of the
Rayleigh-Plateau instability analysis: the RP unstable frequencies
and wavenumbers must satisfy
\eq\{\omega R_0, kR_0 \} \gg \frac{\sigma}{\rho_0 R_0}\,.
\label{validRP}\eeq
Since the most unstable mode dominates the instability we must
guarantee that this condition is verified in the vicinity of the
maximum of the dispersion relation. We find that for
$\frac{\sigma}{\rho_0 R_0}\lesssim 10^{-4}$ this relation is
satisfied, and things get better as $\frac{\sigma}{\rho_0 R_0}$
becomes smaller. In particular this is true in the dispersion
relation plot of Fig. \ref{fig:GLmagnetic}.

Second, the interface between the confined and deconfined phases,
\ie the fluid surface, has a finite thickness of the order $1/T_c$
\cite{Aharony:2005bm}, and therefore the delta-like surface
approximation we used is valid provided that the curvature of the
surface is small with respect to the scale $1/T_c$. So, for plasma
tubes the analysis is good when the boundary radius $R_o$ is
everywhere large when compared with $T_c^{-1}$,
\eq \frac{\sigma}{\rho_0 R_0} \ll 1 \,, \eeq
which is perfectly compatible with (\ref{validRP}).

Finally, we neglected the dependence of the surface tension on the
temperature and other thermodynamic quantities of the fluid. For
consistency we must then demand that on the boundary between the
confined and deconfined phases the temperature of the plasma must
remain everywhere close to the critical temperature $T_c$, \ie
${\cal T}/T_c\sim 1$ at the tube boundary.

As discussed already in great detail in subsection \ref{RPGL}, we
also neglected the dependence of the magnetic susceptibility on the
temperature and on other thermodynamic quantities. In the previous
section we already identified the effects of taking this assumption
and of relaxing it.

Once these conditions are fulfilled, the plasma approximation holds
and can be trusted to study the properties of black holes. The moral
of this analysis is that there is a broad window of parameters for
which the analysis of the Rayleigh-Plateau instability is well
within the required validity regime.

%%%%%%%%%%%%%%%%%%%%%%%%%%%%%%%%%%%%%%%%%%%%%%%%%%%%%%%%%%%%%%%%%%%%%%

\section*{Acknowledgements}

We thank Roberto Emparan for useful discussions.  MC and OJCD are
grateful to the organizers of the workshop ``Quantum Black Holes,
Braneworlds and Holography", Valencia, May 2008; OJCD and DK thank
CERN for hospitality during the programme ``Black Holes: A Landscape
of Theoretical Physics Problems", August-October 2008; OJCD  thanks
the Niels Bohr Institute for hospitality and the organizers of the
workshop ``Mathematical Aspects of General Relativity", Copenhagen,
April 2008, where part of this work was done. MC was supported in
part by DURSI 2005 SGR 00082, MEC FPA 2004-04582-C02 and
FPA-2007-66665-C02, the European Community FP6 program
MRTN-CT-2004-005104, the FWO\,-\,Vlaanderen, project G.0235.05 
and in part by the Federal Office for ScientiÞc, Technical and Cultural Affairs 
through the ÔInteruniversity Attraction Poles Programme Ð Belgian Science 
PolicyÕ P6/11-P. 
 OJCD acknowledges financial support provided by
the European Community through the Intra-European Marie Curie
contract MEIF-CT-2006-038924. This work was partially funded by FCT
through project PTDC/FIS/64175/2006. DK was supported in part by
INFN, MIUR-PRIN contract 20075ATT78, and by the European Community
FP6 program MRTN-CT-2004-005104.

%%%%%%%%%%%%%%%%%%%%%%%%%%%%%%%%%%%%%%%%%%%%%%%%%%%%%%%%%%%%%%%%%%%%%%%%%%%%%%%

\end{document}